 \documentclass{emulateapj}

\newcommand {\Lya}    {Ly$\alpha$}   
\newcommand {\Lyb}    {Ly$\beta$}    

\newcommand {\HI}      {\ion{H}{1}}      
\newcommand {\HeI}   {\ion{He}{1}}   
\newcommand {\HeII}   {\ion{He}{2}}   

\newcommand {\OI}      {\ion{O}{1}}       
\newcommand {\OII}     {\ion{O}{2}}       
\newcommand {\OIII}    {\ion{O}{3}}       
\newcommand {\OIV}    {\ion{O}{4}}       
\newcommand {\OV}    {\ion{O}{5}}        
\newcommand {\OVI}    {\ion{O}{6}}        

\newcommand {\CII}    {\ion{C}{2}}           
\newcommand {\CIII}   {\ion{C}{3}}           
\newcommand {\CIV}    {\ion{C}{4}}          

\newcommand {\NII}     {\ion{N}{2}}
\newcommand {\NIII}     {\ion{N}{3}}
\newcommand {\NIV}     {\ion{N}{4}}
\newcommand {\NV}     {\ion{N}{5}}

\newcommand {\SiIV}   {\ion{Si}{4}}

\newcommand {\SiII}   {\ion{Si}{2}}

\newcommand {\NeV}  {\ion{Ne}{5}}   
\newcommand {\NeVIII}  {\ion{Ne}{8}}   

\newcommand {\FeIII}   {\ion{Fe}{3}} 
\newcommand {\FeII}    {\ion{Fe}{2}}

\newcommand {\NHI}    {$N_{\rm HI}$}

\newcommand {\kms}    {km~s$^{-1}$}

\newcommand {\FUSE}   {{\it FUSE}} 
\newcommand {\HST}    {{\it HST}}
\newcommand{\IUE}   {{\it IUE}}  

\newcommand {\etal}   {et~al.} 

\begin{document}

\title{{\it HST}-COS Observations of AGN. I. Ultraviolet Composite Spectra of the Ionizing Continuum and Emission Lines
\footnote{Based on observations made with the NASA/ESA {\it Hubble Space Telescope}, obtained from the
data archive at the Space Telescope Science Institute. STScI is operated by the Association of Universities for
 Research in Astronomy, Inc. under NASA contract NAS5-26555.}  }  

\author{J. Michael Shull, Matthew Stevans, \& Charles W. Danforth}
\affil{CASA, Department of Astrophysical \& Planetary Sciences, \\
University of Colorado, Boulder, CO 80309}

\email{michael.shull@colorado.edu,  matthew.stevans@colorado.edu, charles.danforth@colorado.edu}  


\begin{abstract} 
The ionizing fluxes from quasars and other active galactic nuclei (AGN) are critical for interpreting the emission-line 
spectra of AGN and for photoionization and heating of the intergalactic medium.   Using ultraviolet spectra from 
the Cosmic Origins Spectrograph (COS) on the {\it Hubble Space Telescope} (\HST), we have directly measured the 
rest-frame ionizing continua and emission lines for 22 AGN.   Over the redshift range $0.026 < z < 1.44$, COS samples the 
Lyman continuum and many far-UV emission lines (\Lya\ $\lambda1216$, \CIV\ $\lambda1549$, \SiIV/\OIV] $\lambda1400$, 
\NV\ $\lambda1240$, \OVI\ $\lambda1035$).   Strong EUV emission lines with 14--22~eV excitation energies 
(\NeVIII\ $\lambda\lambda 770,780$, \NeV\ $\lambda569$,  \OII\ $\lambda834$, \OIII\ $\lambda833, \lambda702$, 
\OIV\ $\lambda788,608,554$, \OV\ $\lambda630$, \NIII\ $\lambda685$) suggest the presence of hot gas in the broad 
emission-line region.   The rest-frame continuum, $F_{\nu} \propto \nu^{\alpha_{\nu}}$, shows a break at 
wavelengths $\lambda < 1000$~\AA, with spectral index $\alpha_{\nu} = -0.68 \pm 0.14$ in the FUV (1200--2000~\AA) 
steepening to $\alpha_{\nu} = -1.41 \pm 0.21$ in the EUV (500--1000~\AA).  The COS EUV index is similar to that of 
radio-quiet AGN in the 2002 \HST/FOS survey ($\alpha_{\nu} = -1.57\pm0.17$).  We see no Lyman edge ($\tau_{\rm HI} < 0.03$) 
or \HeI\ $\lambda584$ emission in the AGN composite.   Our 22 AGN exhibit a substantial range of FUV/EUV spectral 
indices and a correlation with AGN luminosity and redshift,  likely due to observing below
the 1000~\AA\ break. 

\end{abstract} 



\section{INTRODUCTION}

Active galactic nuclei (AGN) hold a place of special importance in astrophysics, as examples 
of processes of high-energy emission and gas accretion around super-massive black holes and 
for their impact on surrounding gas in the nuclei of galaxies and throughout the intergalactic 
medium (IGM).  Their rest-frame Lyman continuum (LyC) emission ($\lambda \leq 912$~\AA)
is likely to be the dominant source of metagalactic ionizing radiation over most of the age of the universe.  
The shape of their ionizing continua is critical for interpreting their broad emission-line spectra, and provides 
the ionization corrections required to derive abundances of the common (\CIV, \SiIV, \OVI) metal-line systems 
in quasar absorption spectra (Schaye \etal\ 2003; Aguirre \etal\ 2004; Danforth \& Shull 2008).
The AGN ionizing continua at 1 ryd and 4 ryd are also important for interpreting fluctuations in 
the ratio of the \Lya\ absorbers of \HI\ and \HeII\ (Fardal \etal\ 1998; Fechner \etal\ 2006;
Shull \etal\ 2004, 2010).  
 
The nuclei of many galaxies exhibit emission lines characteristic of photoionization, both from
hot stars and non-thermal continua.   In a subset, the emission lines exhibit a range of ionization
states and line widths uncharacteristic of normal or starburst galaxies.  These AGN include objects 
classified as Seyfert galaxies, quasars (quasi-stellar radio galaxies), and QSOs (quasi-stellar objects).  
Their emission lines are believed to be produced by a photoionizing continuum extending from the
extreme-ultraviolet to the X-ray,  which may be produced by an accretion disk around the black hole and
by relativistic electrons, involving synchrotron emission and Compton scattering.  The ionizing spectrum 
extends to energies well beyond the hydrogen edge ($h\nu \geq 1$~ryd), in order to produce strong 
emission lines (Krolik \& Kallman 1988) from high ions such as \SiIV\  (2.46 ryd), \CIV\ (3.51 ryd), 
\HeII\ (4.00 ryd),  \NV\ (5.69 ryd), and \OVI\ (8.37 ryd).  

Although AGN are observed over a wide variety of luminosities and physical circumstances 
(mass, viewing angle, accretion rate), they display a consistent {\it mean} spectral energy distribution 
(SED) in the UV-optical.  Individual sources exhibit considerable variations in the far ultraviolet
(FUV, 1000--2000~\AA) and extreme ultraviolet (EUV, $\lambda < 912$~\AA), with observed flux 
distributions fitted to power laws, $F_{\lambda}  \propto \lambda^{\alpha_{\lambda}}$ and
$F_{\nu} \propto \nu^{\alpha_{\nu}}$, where $\alpha_{\nu} =  -(2 + \alpha_{\lambda})$.  
The FUV and EUV spectra afford a glimpse of the inner workings of the black hole accretion disk and the
broad emission-line region (BELR).  Ground-based surveys of AGN at $z > 2$ have characterized the rest-frame 
ultraviolet, longward of the AGN \Lya\ emission line and free of intergalactic hydrogen absorption.  Composite 
UV spectra of AGN have been compiled by many groups (Francis \etal\ 1991;  Schneider \etal\ 1991; 
Carballo \etal\ 1999;  Brotherton \etal\ 2001; Vanden Berk \etal\ 2001).    All find a wide range of spectral indices, 
steepening toward high frequencies, with possible dependence on sample redshift and luminosity range.  

The best direct probe of the rest-frame FUV and EUV continua comes from ultraviolet (space-borne) observations 
of AGN at $z \leq 1.5$.  Although the rest-frame continuum shortward of 1216~\AA\ is accessible from the
ground toward higher-redshift AGN, it is often complicated by intergalactic \Lya\ absorption (Fan \etal\  2001).  
Thus, ultraviolet spectrographs aboard the {\it International Ultraviolet Explorer} (\IUE), 
{\it Far Ultraviolet Spectroscopic Explorer} (\FUSE), and {\it Hubble Space Telescope} (\HST) have provided 
access to the AGN continuum below \Lya\ in a number of bright, low-$z$ targets.  Composite spectra of these AGN were 
created from \HST/FOS and \HST/GHRS observations (Zheng \etal\ 1997; Telfer \etal\ 2002) and from \FUSE\ (Scott \etal\ 2004).  
For 184 QSOs at $z > 0.33$ studied with \HST/FOS, Telfer \etal\ (2002) fitted the composite continuum (500--1200~\AA) 
as a power law with spectral index $\alpha_{\nu}  = -1.76\pm0.12$.  For a sub-sample of 39 radio-quiet QSOs, they found 
$\alpha_{\nu}  = -1.57\pm0.17$.  In contrast, the \FUSE\ survey of 85 AGN at $z \leq 0.67$ (Scott \etal\  2004) found a harder 
composite spectrum with $\alpha_{\nu} = -0.56^{+0.38}_{-0.28}$ and suggested that the difference may reflect a luminosity 
dependence through the well-known anti-correlation (Baldwin effect) between emission-line strength and AGN luminosity.   
The difference could also arise from the small numbers of targets observed by the different instruments;  fewer than 
10 AGN observations cover the spectral range $450~{\rm \AA} \la \lambda \la 600$~\AA.  In all of these studies, a crucial
issue is the placement of the EUV continuum, which sits below strong emission lines such as \NeVIII, \NeV, \OIII, \OIV, \OV, 
and \OVI.  

Figure 1 shows current estimates of the metagalactic ionizing continuum, interpolated between far-UV and X-ray 
energies.   At the Lyman limit (1~ryd), we adopt a specific intensity
$I_{\nu} \approx 2 \times 10^{-23}~{\rm erg~cm}^{-2}~{\rm s}^{-1}~{\rm Hz}^{-1}~{\rm sr}^{-1}$ 
(Shull \etal\ 1999; Haardt \& Madau 2012) and use the two ``bow-tie" extrapolations derived in the composite AGN 
spectra from \HST\ and \FUSE.    In the X-ray band (0.5--2~keV), we use background estimates from {\it ROSAT} measurements 
of AGN in the Lockman Hole (Hasinger \etal\ 1993) normalized at 1 keV with energy flux $F_E \propto E^{-0.96 \pm 0.11}$.   
The ionizing intensity in the mid-EUV remains uncertain, depending on how one interpolates between the measured 
fluxes at 1.0--1.5 ryd and 0.5--2  keV.

In this paper, we chose an initial COS sample of 22 AGN (non-blazar) at $z \leq 1.5$,  available in  2011 January  and 
selected to have S/N sufficient to directly measure their rest-frame UV continua and emission lines.   
With minimal contamination by \Lya-forest absorption, we develop a composite AGN spectrum in the FUV and EUV for targets 
at redshifts between $z = 0.026$ and $z = 1.44$.  We identify the prominent AGN emission lines and line-free portions of the 
spectrum and fit the underlying continua below the emission lines, excluding interstellar and intergalactic absorption lines.  
In Section 2 we describe the COS data reduction, our techniques for flux-alignment and patching \HST/COS spectra 
with earlier UV spectra taken by \IUE\  and \FUSE.    In Section 3 we describe our results on the 
individual and composite spectra.  Section 4 presents our conclusions and their implications.


\begin{figure}
\epsscale{1.2}
\plotone{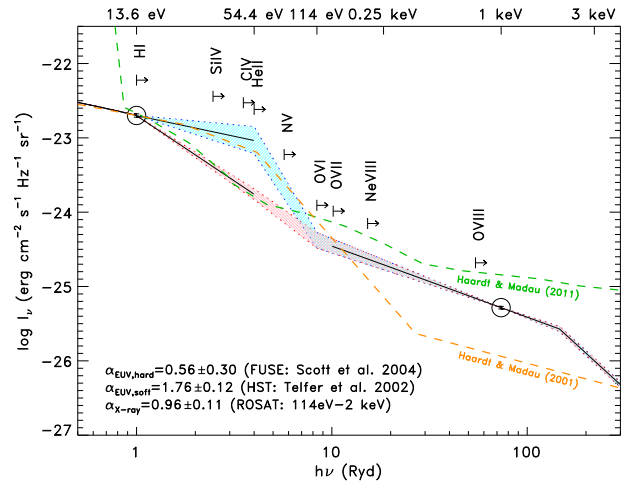}   
\caption{Two possible model spectra of AGN ionizing radiation (EUV to X-ray), normalized at 1 ryd with
   specific intensity $I_0 = 2 \times 10^{-23}$ erg~cm$^{-2}$~s$^{-1}$ Hz$^{-1}$ sr$^{-1}$ and two different flux
   distributions, $F_{\nu} \propto \nu^{\alpha_{\nu}}$, between 1 and 4 ryd.  Previous composite AGN spectra are shown 
   in red (\HST) and light blue (\FUSE), with slopes $\alpha_{\nu} = -1.76 \pm 0.12$ (Telfer \etal\ 2002) and 
   $\alpha_{\nu} = -0.56^{+0.38}_{-0.28}$ (Scott \etal\ 2004).   Soft X-ray data (0.5--2 keV) are from {\it ROSAT}
   observations of AGN in the Lockman Hole (Hasinger \etal\ 1993;  Hasinger 1994) 
   normalized at $E = 1$ keV (73.53 ryd) with specific energy flux $F_E \propto E^{-0.96 \pm 0.11}$.   
   Ionization potentials of \HI\ and \HeII\  and energies required to produce key metal ions are shown as arrows.
   Metagalactic backgrounds from Haardt \& Madau (2001, 2012) are also shown.} 
\end{figure}


\section{OBSERVATIONS OF ULTRAVIOLET SPECTA OF AGN}  

\subsection{Data Acquisition and Reduction Methods} 

Table~1 lists the relevant COS observational parameters of  22 AGN (non-blazar)  targets,  selected in 2011 January 
as the best available high-S/N spectra of AGN.   This initial sample was not intended to be archivally complete, as we 
discuss at the end of Section 2.  Most of the low-redshift AGN are Seyfert galaxies.  Several AGN exhibit narrow intrinsic 
absorption lines (e.g., Mrk\,817, Mrk\,290, Mrk\,1513, PG\,1115+407).  With the spectral resolution and high throughput 
of COS, we are able to fit out all these absorbers and measure the underlying continuum.  
All targets were observed in both the G130M  (1135--1460~\AA) and G160M (1390--1795~\AA)  gratings which
provide medium resolution ($R \equiv \lambda/\Delta \lambda \approx 18,000$ or $\Delta v  \approx 17$ \kms).   
The COS instrument and data acquisition are described by Osterman \etal\ (2011) and Green \etal\ (2012).  
The NUV-imaging target acquisitions were performed with the MIRRORA/PSA mode.  The COS primary science 
aperture ($\sim2.5''$ diameter) yields good centering to maximize the COS throughput and spectral resolving power.

Each sight line was observed with four or more exposures in both the G130M and G160M gratings.  The central 
wavelengths were shifted between exposures to dither instrumental features in wavelength space and to provide a 
continuous spectrum over the entire FUV spectral range.  Individual exposures for each sight line were retrieved from 
the Multi-mission Archive for Space Telescope (MAST) processed with the default {\sc CalCOS} reduction pipeline.  
Alignment and coaddition of the individual exposures was carried out using IDL routines developed by the COS GTO 
team specifically for FUV spectra (IDL routines are available at
{\small {\tt  \url{http://casa.colorado.edu/$\sim$danforth/costools.html}}}.)
Danforth et al.\ (2010) describe the process in detail;  the only significant difference is our omission of the pseudo flatfielding 
procedure from that paper.  A similar procedure is now included in the standard {\sc CalCOS} reduction.   
Data coalignment was achieved with strong interstellar features in each exposure, cross-correlated and 
interpolated onto a common wavelength scale.  The wavelength shifts were typically on the order of a resolution 
element ($\sim 0.1$~\AA) or less.  The coadded flux at each wavelength was taken to be the 
exposure-weighted mean of flux in each aligned exposure.  Exposure times at the edges of detector segments 
were adjusted downward to underweight these regions.

\subsection{Fitting and Patching the UV Spectra}

Here, we discuss how we aligned the fluxes, dealt with detector gaps, dereddening, Lyman-limit systems, cosmological factors in 
rest-frame wavelength and specific flux, and other technical issues.    We then provide details of how we constructed 
the composite spectra, with brief discussion of the order of adding spectra and choosing wavelength-pivot points.  
To combine the non-simultaneous observations from \FUSE, \HST/COS, and \IUE, we scale the flux within the regions of spectral 
overlap to agreement.  The \FUSE\ and COS spectra overlap by $\sim45$~\AA, those of COS and \IUE\ by $\sim645$~\AA, and those 
of \IUE-SWP (short-wavelength-prime channel) and \IUE-LWR (long-wavelength-redundant channel) by $\sim125$~\AA.  When the 
overlapping regions contain uncontaminated continuum flux, we use only these regions to scale the data. The \FUSE\ and \IUE\ data 
are scaled to \HST/COS because, to our knowledge, the flux calibration of the \HST/COS spectra is more reliable. The \IUE-SWP and
\IUE-LWR fluxes appear consistent and are not scaled to each other.  Scaling factors for the \IUE-to-COS spectra range from 0.64 to 
2.07 with an average of 1.13,  and for the  \FUSE-to-COS data from 0.35 to 1.38 with an average of 0.90.   

Source variability could affect the absolute flux and shape of AGN SEDs in the ultraviolet.  For example, IUE-AGN 
variability campaigns found that the UV continua  hardened as the flux increased for the Seyfert galaxies NGC\,4151 
(Crenshaw \etal\ 1996; Edelson \etal\ 1996), NGC~5548 (Clavel \etal\ 1991), and NGC~3783 (Reichert \etal\ 1994).  For Fairall~9 
(Rodriguez-Pascual \etal\ 1997) and NGC~3516 (Edelson \etal\ 2000) there was no significant change in the relative FUV flux.
Determining the actual shape of the underlying UV continuum was challenging in these studies, owing to spectral dilution by the
\FeII\  pseudo-continuum (NUV) and difficulty finding line-free continuum windows.  The combination of recent and archival 
data could make our SEDs vulnerable to this effect.    However, because we primarily use coeval COS data to measure
the slope of the continuum,  our scaling should not greatly influence the result.  

As noted earlier, several of our Seyferts and AGN show outflows and intrinsic absorption common in Seyfert galaxies and 
QSOs (Crenshaw \etal\ 2003).  In the seven low-$z$ targets ($z \leq 0.154$) the \CIV\ line provides the best indicator of such
outflows (Mrk~817, Mrk~290, NGC~1513, PG~1115+407). At higher redshift, \CIV\ shifts out of the COS bands, but one can
use \OVI\ $\lambda1035$ as an indicator (e.g., HE~0153-4520, HE~0226-4110, HE~1102+3441).  For the highest-redshift AGN, 
one can use \NeVIII\ $\lambda\lambda$770, 780 to probe outflows.  For our continuum fitting, we are able to
fit out all the absorbers:  intrinsic systems as well as interstellar and intergalactic lines.  This capability is made 
possible by the high-S/N and high-resolution of COS.  We also mask out Galactic \Lya\  absorption (1215.67~\AA) and the 
geocoronal emission line of \OI\ $\lambda1304$ in every spectrum.

We correct the spectra for Galactic reddening, using the empirical mean extinction curve of Fitzpatrick (1999)  with a ratio of 
total-to-selective extinction $R_v = A_V / E(B-V) = 3.1$ and color excesses $E(B-V)$ from 
NED\footnote{NASA/IPAC Extragalactic Database (NED) is operated by the Jet Propulsion Laboratory, California Institute of 
Technology, under contract with the National Aeronautics and Space Administration, {\tt \url{http://nedwww.ipac.caltech.edu.}} }
based on dust mapping by Schlegel \etal\ (1998).  For our purposes, this extinction curve is extrapolated down to 1000~\AA.  
If we use the extinction curve of Cardelli \etal\ (1989), the spectral index changes by  only $\Delta \alpha \approx 0.02$.
We inspect the spectra for absorption from  Lyman limit  systems (LLS) and partial LLS  by identifying many Lyman lines that are 
blueshifted from Galactic absorption. Table 2 shows the strengths (column densities \NHI) and redshifts of 17 LLS found
toward 8 AGN in our sample.  We account for Lyman continuum absorption by correcting for the $\nu^{-3}$  opacity shortward of each 
Lyman edge using the optical depth, $\ln (F_+ /  F_-)$, where $F_-$ is the median flux of the pixels just blueward of the break and 
$F_+$ is the median flux of the pixels in selected windows longward of the break.   We have used the restored continua in all
cases except for the strong LLS toward SBS\,1108+560 ($\log N_{\rm HI} = 17.78$), for which we ignore the spectrum 
below 755~\AA\ (rest frame).    Finally, we apply redshift corrections to shift each spectrum to the AGN rest frame by dividing the 
wavelengths by  $(1+z)$.

We use the IDL routine MPFIT\footnote{ {\tt \url{http://nedwww.ipac.caltech.edu}}  (see also Markwardt 2009) } to fit the continuum of each 
spectrum with a power law of form, $F_{\lambda} \propto  \lambda^{\alpha_{\lambda}}$, with spectral indices, $\alpha_{\lambda}$, 
listed in Table~1.  Because of the strong emission lines commonly seen in AGN spectra, finding windows of completely uncontaminated 
continuum can be difficult.  We focus on wavelength regions with the lowest relative flux (local minima) that appear to be free 
of emission lines.  We avoid the prominent emission from \Lya\ $\lambda1216$ and the doublets of
 \CIV\ $\lambda\lambda1548,1551$, \NV\ $\lambda\lambda1239,1243$,  \OVI\ $\lambda\lambda1032,1038$, and \SiIV\  
 $\lambda\lambda1394,1403$.   The \SiIV\ lines are typically blended with \OIV] $\lambda1405$, and the \OVI\ lines with 
 \Lyb\ $\lambda 1025$.  We also detect a number of weaker lines,  known from atomic data and clearly seen in Figure 2 in
 high-S/N spectra of Mrk~817 and Mrk~1513.   These include line blends of \OI\ $\lambda1304$ and \SiII\ $\lambda1308$, 
 \HeII\ $\lambda1640$,  \OIII] $\lambda1605$, \CII\ $\lambda1335$, \NII\ $\lambda1085$, \FeIII\ $\lambda1126$, and \NIV] $\lambda1486$.  
These features appear at the expected wavelengths, although in some cases residual flux from these blends appears as a pseudo-continuum.   
The high resolution of  ($\sim20$~\kms) and sensitivity of COS (S/N listed in Table 1) allow us to exclude many absorption lines
from the Galactic interstellar medium and IGM.  The most common regions of  (rest-frame, line-free) continuum (listed here in \AA) include:  
660--670, 715--735, 860--880, 1090--1105, 1140--1155, 1280--1290, 1315--1325, and 1440--1465.



\begin{deluxetable*}{lcllllll}
\tabletypesize{\footnotesize}
\tablecaption{COS Observations (22 AGN Targets) }
\tablecolumns{8}
\tablewidth{0pt}
\tablehead{
  \colhead{AGN\tablenotemark{a}}                                             &
  \colhead{AGN\tablenotemark{a}}                                              &
  \colhead{$z$\tablenotemark{a}}                                                &
  \colhead{$\alpha_{\lambda}$}                                                   &
  \colhead{$F_0$\tablenotemark{a}}                                           &       
  \colhead{$F_{\lambda}$\tablenotemark{a}}                            &
  \colhead{$\log (\lambda L_{\lambda})$\tablenotemark{a}}  &  
  \colhead{(S/N)$_{\rm res}$\tablenotemark{a}}  \\
  \colhead{Target} & \colhead{Type}    &   &   &   \colhead{(1100~\AA)}  &   \colhead{(1300~\AA)}  & (1100~\AA)  &   
            }
 \startdata 
    
Mrk 335                & Sy~1      &   0.025785     &   -1.35      &  $5.68$   & $4.0$         &    43.98   &  71, 18    \\  
Mrk 290                & Sy~1.5   &   0.029577     &   -1.05      &  $1.80$   & $1.9$         &    43.60   &  63, 26   \\  
Mrk 817                & Sy~1.5   &   0.031455     &   -1.49      &  $9.51$   & $10.0$       &    44.38   &  73, 36    \\  
PG 1011-040      & Sy~1.2   &   0.058314      &    -1.75     &  $2.75$   & $3.2$         &    44.39   & 38, 21     \\
Mrk 1513              & Sy~1.5  &   0.062977      &    -1.09     &  $2.73$    & $5.7$         &    44.46   & 36, 8      \\
Mrk 876                & Sy~1      &   0.129            &    -1.19     &  $4.90$    & $3.8$         &    45.38   & 72, 55    \\     
PG 1115+407     & Sy~1      &  0.1546           &    -1.42     &  $1.07$    & $1.0$         &    44.89   & 27, 20    \\  
                               &                &                         &                  &                    &                    &                 &                 \\  
HE 0153-4520    & QSO      &  0.451             &   -0.80      &   $1.36$    & $1.4$          &    46.06  & 33, 29    \\  
PG 1259+593     & Sy~1     &  0.4778           &   -1.09      &  $1.18$     & $1.4$          &    46.06  & 43, 35   \\
HE 0226-4110    & Sy~1     &  0.493368      &    -1.47     &  $1.73$     & $2.0$          &    46.26  & 49, 38   \\  
HS 1102+3441   & QSO      &  0.5088          &   -1.26      &   $0.255$  & $0.30$        &    45.46  & 25, 27     \\  
HE 0238-1904     & QSO     & 0.631             &   -0.42      &   $1.48$     & $1.3$          &    46.45  & 38, 26   \\  
3C 263                  & Sy~1.2  & 0.6460          &   -1.03      &   $0.693$   & $1.0$          &    46.14  & 47, 38    \\
SBS1108+560     & QSO     & 0.7666           &   -0.84      &  $0.422$    & $0.013^*$  &    46.11  &  4, 16 \\  
                               &                &                        &                  &                     &                     &                 &              \\  
SBS 1122+594    & QSO      & 0.852            &  -0.62       &  $0.231$     &  $0.24$       &    45.96  &  18, 15   \\
FBQS J0751+2919 & QSO  &  0.9149        &  -0.51       &  $0.731$     & $0.65$        &     46.54  & 40, 33   \\  
PG1407+265        &  QSO    & 0.95               &   -1.56     &  $0.671$     &  $1.1$         &     46.55  & 66, 36  \\  
PG1148+549        & QSO     & 0.9754          &   -0.53     &  $0.375$     & $0.47$        &     46.32  &  41, 33  \\  
HE 0439-5254      & QSO     &  1.053	           &   -1.31     &  $0.201$     & $0.39$        &     46.13  &  23, 17  \\
PG 1206+459        & QSO    & 1.1625          &   -0.28     &   $0.453$     & $0.36$       &     46.59  &  30, 33 \\  
PG 1338+416         & QSO    & 1.214            &   -1.17     &  $0.0874$   & $0.17$       &     45.93  &  26, 23  \\  
Q0232-042	      & QSO    &  1.437368    &   -0.67     &  $0.196$     & $0.22$        &     46.46  & 24, 17  \\  

\enddata

\tablenotetext{a}{AGN targets, types, redshifts, fluxes, spectral indices, luminosities, and S/N ratios.  
   All fluxes in units of $10^{-14}$~erg~cm$^{-2}$~s$^{-1}$~{\AA}$^{-1}$. Rest-frame, dereddened spectral 
   distributions are fitted to power laws, $F_{\lambda} = F_0 (\lambda / 1100~{\rm \AA})^{\alpha_{\lambda}}$.   
   Wavelength index $\alpha_{\lambda}$ corresponds to frequency index 
   $\alpha_{\nu} = -[2 + \alpha_{\lambda}]$.  The 22 AGN are listed in three redshift groups: 
   low ($0.026 \leq z \leq 0.154$ with $\langle \alpha_{\lambda} \rangle = -1.33$);
   intermediate ($0.45 \leq z \leq 0.77$ with $\langle \alpha_{\lambda} \rangle = -0.99$);
   high ($0.85 \leq z \leq 1.44$ with $\langle \alpha_{\lambda} \rangle = -0.83$).
   The eight columns show:  
   (1)  AGN target;  (2) AGN type;  (3) AGN redshift;  (4) Fitted spectral index $\alpha_{\lambda}$;  
   (5) Rest-frame flux normalization $F_0$ at 1100~\AA; (6) Observed flux $F_{\lambda}$ at 1300~\AA;
   (7) Band luminosity, $\lambda L_{\lambda}$ at 1100~\AA\ (in erg~s$^{-1}$); 
   (8) signal-to-noise at 1250~\AA\ and 1550~\AA\ for data with G130M (1132--1460~\AA) and 
   G160M (1394--1798~\AA) gratings, respectively.  
   Flux at 1300~\AA\  for SBS~1108+560 (noted with $^*$) is low,  owing to LyC absorption 
   ($\lambda < 1334$~\AA) from a Lyman limit system at $z = 0.46335$.     }

\end{deluxetable*}


\newpage

\subsection{Constructing and Fitting the Composite Spectrum}
	
The spectra are resampled to wavelength bins of 0.1~\AA, using the same procedure as in Telfer \etal\ (2002).   Before resampling,
we discard any pixels with S/N $< 1$.    The flux corresponding to each new wavelength bin is the mean of the flux in the old pixels 
that overlap the new bin, weighted by the extent of overlap. The error arrays associated with the resampled spectra are determined 
using a weighting method similar to the flux rebinning.  The resampled fluxes are calculated using 
 \begin{equation}
    F_r  =   \frac { \sum_i  F_i  (\delta \lambda'_i ) }  { \sum_i  (\delta \lambda'_i) }   \; , 
\end{equation}
and their errors are the same as presented in Telfer \etal\ (2002).  
Here, $F_r$ is the flux of the rebinned pixel, and $F_i$ are the fluxes of pixels in the original  spectrum.  The widths $(\delta \lambda_i)$ are the 
sizes of the original wavelength bins, and the $(\delta \lambda'_i)$ are the overlap of the old pixels with the new bin.  
Our initial composite spectrum used the bootstrap technique from Telfer \etal\ (2002) to combine the resampled spectra.  We began with the 
spectrum of HE~0238-1904, which at $z=0.631$ lies at the median redshift of our sample.  We defined it as our first partially formed composite 
spectrum and renormalized for convenience. We then included the remaining spectra in sorted order, alternating between higher-redshift spectra 
and lower-redshift spectra, after normalizing the continuum-like regions of each spectrum to the continuum of the partially formed 
composite spectrum. 

We then explored ways of improving the composite.  Ideally we would normalize all the spectra in the sample to each other at one common
continuum window;  the wide range of redshifts precludes such an approach.  Instead, we find the rest-frame continuum window common to
the most spectra (860--880~\AA\ is shared by 12 of 22 spectra).  We designate the spectra that share this continuum window as Group 1 and 
normalize those spectra within that wavelength region. A partially formed composite spectrum is produced by calculating the mean flux in each 
0.1~\AA\ bin and renormalizing for convenience. The remaining spectra are separated into two other groups:
Group 2 contains AGN at smaller redshifts than Group~1, and Group~3 are AGN at larger redshifts.  We compile the spectra of 
Groups 2 and 3 with the Group 1 composite in the following steps:   
(1) normalize to the partially formed composite;  (2) calculate a new  partially formed composite;
(3)  renormalize the new partially formed composite;  (4) repeat until all spectra in Groups 2 and 3 are compiled in the final composite.
We normalize the spectra of Groups 2 and 3 to the partially formed composite by finding the weighted-mean normalization constant from 
multiple continuum windows, calculated using Equation (4) of Telfer et al (2002).  Because the Group 1 composite represents the central 
region of the final composite, the order of adding Group-2 and Group-3 spectra to the composite does not matter.  In practice, we added 
Group 2 spectra in order of decreasing redshift and Group-3 spectra in order of increasing redshift. To prevent narrow absorption features 
from affecting the normalization within the continuum windows, we normalize using spline fits to the individual spectra and use a partially 
formed composite of the spline fits. 

Our final AGN composite spectrum uses spline fits to the individual spectra rather than the raw data themselves.  This has the advantage of 
interpolating over narrow absorption features, which would otherwise complicate the coaddition of higher-redshift spectra.  We did not 
include the restored continuum data from SBS~1108+560 at $\lambda < 770$~\AA, owing to the low S/N from a strong LLS.  
We fit continua to each of the data sets, using a semi-automated line identification and spline-fitting technique as follows.  First, the
spectra are split into 5--10 \AA\ segments.  Continuum pixels within each segment are identified as those for which the S/N 
(per pixel flux/error) vector is less than $1.5 \sigma$  below the S/N in the segment.  Thus, absorption lines are excluded,  
as are regions of increased noise.  The process is iterated until minimal change occurs between one iteration and the next.  The continuum 
pixels in a particular bin are then set, and the median continuum flux node is recorded. A spline function is fitted between continuum nodes.  
We check the continuum fits manually, and the continuum region identifications are adjusted as needed.  The continuum identification 
and spline-fitting processes work reasonably well for smoothly varying data, but they were augmented with piecewise-continuous 
Legendre polynomial fits in a few cases.  In particular, spline fits perform poorly in regions of sharp spectral curvature, such as the 
Galactic Ly$\alpha$ trough and at the peaks of cuspy emission lines.  More details on the process are given elsewhere 
(C. W. Danforth \etal\ 2012,  in preparation).

We now discuss sources of random and systematic uncertainty in the composite spectral indices.  In our process, we fit two power laws to 
the spline composite spectrum, $\alpha_{\rm FUV}$ and $\alpha_{\rm EUV}$,  and match them at a break wavelength 
$\lambda \approx 1000\pm50$~\AA.  Although this break is apparent in the composite, its presence is less clear in the individual spectra.
Because of the AGN redshift distribution of our sample ($z < 0.16$ and $z > 0.45$) the rest-frame spectra lie either above or below 
the 1000~\AA\ break.  To quantify the uncertainty in the fitting of the composite spectrum, we explore sources of uncertainty described by 
Scott \etal\ (2004), including the effects of cosmic variance in the shape of AGN SEDs,  Galactic extinction, and formal statistical fitting errors.  
We do not include the effects of intrinsic absorbers, \Lya\ forest, or interstellar medium;  these absorption lines are easily removed.  
The largest source of uncertainty comes from the natural variation in the slope of the contributing spectra. We estimate this uncertainty 
by selecting 1000 bootstrap samples with replacement from our sample of 22 AGN spectra.   The standard deviations of the resulting spectral 
slope distributions are $\sigma = 0.21$ for $\alpha_{\rm EUV}$ and $\sigma = 0.14$ for $\alpha_{\rm FUV}$.   These turn out to be the 
dominant sources of uncertainty.


\begin{deluxetable}{lllc}
\tabletypesize{\footnotesize}
\tablecaption{Lyman Limit and Partial Limit Systems  }
\tablecolumns{4}
\tablewidth{0pt}
\tablehead{
\colhead{AGN Target}   &   \colhead{$z_{\rm LLS}$\tablenotemark{a}}   &   \colhead{log~N$_{\rm HI}$\tablenotemark{a}}   
    & \colhead{$b$\tablenotemark{a} }  \\
     &  \colhead{(redshift)}    &   \colhead{(\NHI\ in cm$^{-2}$)}    &   \colhead{ (km~s$^{-1}$) }  
 }
\startdata
HE 0439-5254      & 0.61504    & 16.15     &   32         \\
                                & 0.61565     & 15.6       & \nodata  \\
SBS 1122+594    & 0.55743     & 15.9        & 22           \\
                                & 0.55810     & 16.25     & 15            \\
                                & 0.5584       & 15.9        & 20            \\
                                & 0.67822     & 16.3       & 18             \\
SBS 1108+560    & 0.463          & 15.8        & \nodata     \\
                                & 0.46335     & 17.78     & 20              \\
FBQS0751+2919 & 0.82915    & 16.1        & 30              \\
PG 1407+265       & 0.57497    & 15.6        & 28              \\
                                & 0.59954     & 15.9       & 20               \\
                                & 0.6827       & 16.35     & 33               \\
PG 1206+459      & 0.92707     & 16.4       & \nodata      \\
                               & 0.9276       & 17.1        & 30                \\
PG 1338+416      & 0.68615     & 16.7       & 20                 \\
                               & 0.68645     & 15.5        & \nodata       \\
Q 0232-042          & 0.7389        & 16.8        & 35              \\
\enddata

\tablenotetext{a}{For these 8 AGN sight lines, we found 17 Lyman limit or partial Lyman limit systems.
We list their redshifts, H~I column densities, and doppler parameters ($b$) derived by fitting \Lya\
and higher Lyman series absorbers. }  

\end{deluxetable}


We investigated uncertainties that arise from UV extinction corrections from two quantities: color excess $E(B-V)$ and ratio of total to 
selective extinction $R_V$.  We alter the measured $E(B-V)$ by $\pm16$\% ($1\, \sigma$) as reported by Schlegel \etal\ (1998).   
We deredden the individual spectra with $E(B-V)$ multiplied by 1.16 or 0.84, compile the spectra into a 
composite, and fit the continua.  We find that $\alpha_{\rm EUV}$ changes by ($+0.025, -0.021)$ while $\alpha_{\rm FUV}$  changes by 
$(+0.045, -0.043)$.    Next, we estimate the sensitivity to deviations from the canonical value $R_V = 3.1$ by dereddening the 
individual spectra with $R_V = 2.8$ and $R_V = 4.0$ and compiling the spectra into composites.  
We find that $\alpha_{\rm EUV}$ changes by  $(+0.017, -0.044)$ and $\alpha_{\rm FUV}$ by $(+0.026, -0.076)$.
The formal statistical errors for the spectral indices are negligible ($<0.0015$),  owing to the high S/N ratio of our composite spectra, and
we do not include them in the final quoted uncertainties.  We add the random uncertainties of cosmic variance with the systematic 
effects of correcting for extinction in quadrature and estimate the total uncertainties to be $\pm 0.21$ for $\alpha_{\rm EUV}$ and 
$\pm 0.15$ for $\alpha_{\rm FUV}$.

Obtaining a larger AGN sample will be helpful, particularly for the shorter wavelengths ($\lambda < 900$~\AA) for which only eight
AGN contribute.   In the future, we plan to create a composite AGN spectrum between 300 and 1800 \AA, using \HST/COS archival spectra
of AGN with a variety of types and luminosities.   The order-of-magnitude sensitivity increase offered by COS over previous spectrographs 
has increased the number of targets available for far-UV spectroscopy to over 270 AGN.  These archival spectra will provide EUV coverage 
down to 500~\AA\  (with $\sim$40 AGN) and to 912~\AA\ (with $\sim$100 AGN).  
For even shorter rest-frame wavelengths (304--500~\AA) we can use the sample of quasars (Shull \etal\ 2010; Worseck \etal\ 2011; 
Syphers \etal\ 2011) observed to study the \HeII\  reionization epoch.   These ``\HeII\ quasars'' are the rare subset of $z \sim 3$ quasars
that are FUV-bright because of low hydrogen column density.  There are now 31 such FUV-bright quasars observed with COS,  of 
which 27 have flux extending down to 304~\AA.  Below this wavelength, intergalactic \HeII\  absorbs most or all of the flux.  While most of
 these spectra have low S/N, when combined they will extend the composite EUV spectrum out to 3 ryd in energy.    For comparison, 
 Telfer \etal\ (2002) had only one quasar extending to 304~\AA\  and Scott \etal\ (2004) had none.


\begin{figure*}
\epsscale{1.2}
   \plotone{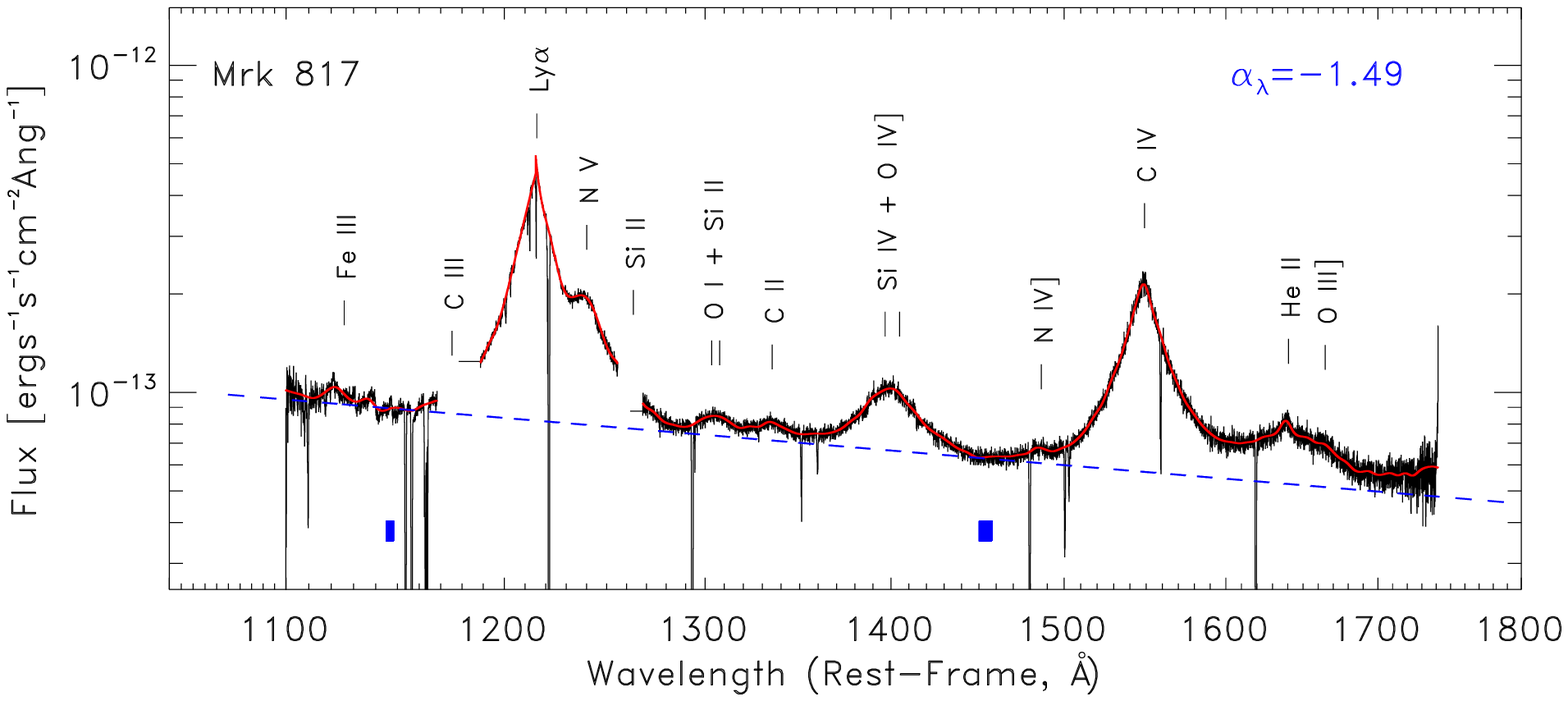}
   \plotone{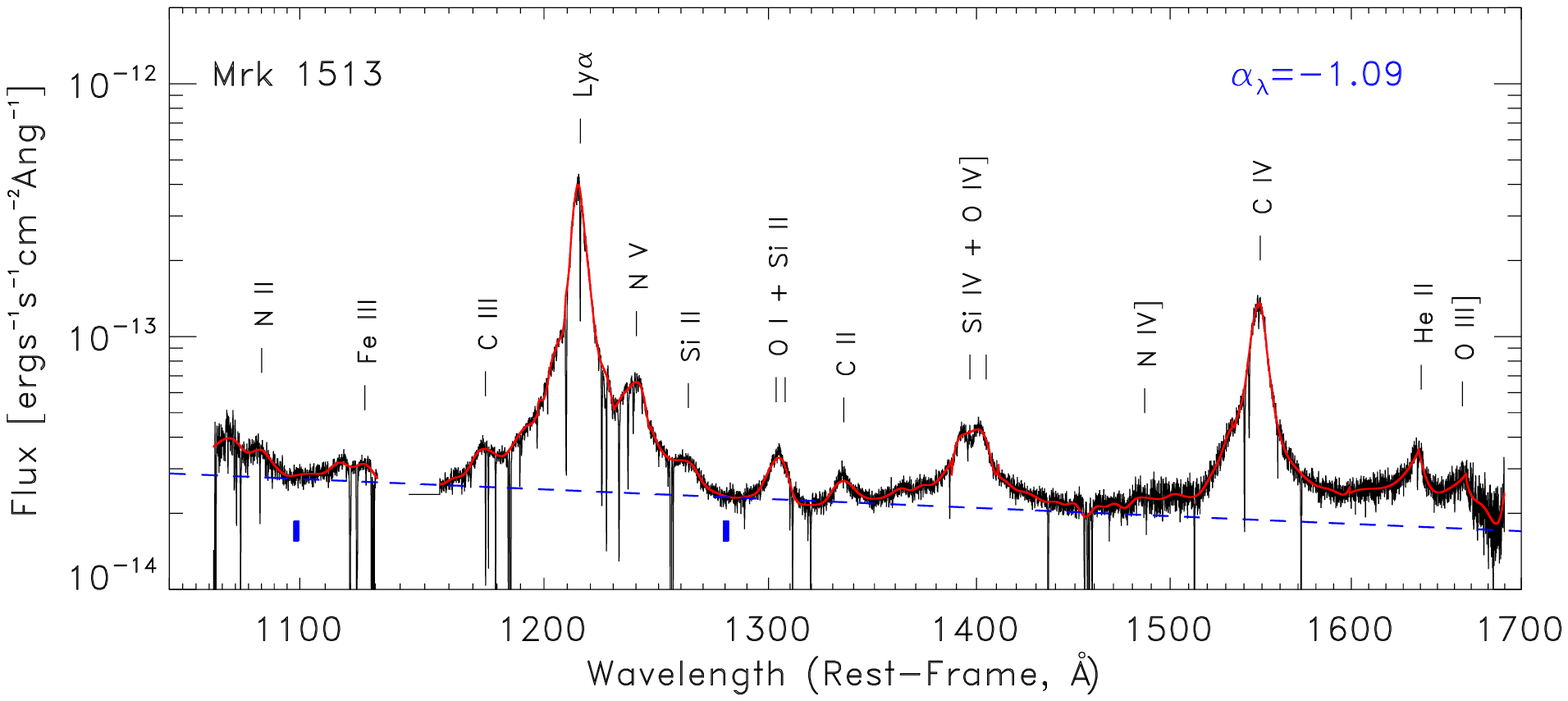}
\caption{COS spectra of Mrk 817 and Mrk 1513, binned to a COS resolution element ($\sim0.08$~\AA) by smoothing 
over 7 pixels and binning by 3 pixels.  Red line is spline fit.  We label prominent emission lines (Table 4), line-free
continua (blue boxes), and excise gaps from Galactic \Lya\ absorption.  Fitted far-UV spectral indices,  $\alpha_{\lambda}$, 
are in top right corners.  
} 
\end{figure*}


\section{RESULTS}
   
The results of our AGN spectral fits and UV/EUV composite spectra are displayed in Figures 2--7.   
Table 1 provides information about the 22 individual AGN, the power-law fits to their \HST/COS spectra, 
and their band-luminosities, $\lambda L_{\lambda}$, at 1100~\AA, given by
\begin{eqnarray}
   \lambda L_{\lambda} &=& (1.32\times10^{43}~{\rm erg~s}^{-1}) \left[ \frac {d_L}{100~{\rm Mpc}} \right]^2  \nonumber \\
         & \times  &    \left[ \frac {F_{\lambda}} {10^{-14}~{\rm erg~cm}^{-2}~{\rm s}^{-1}~{\rm \AA}^{-1}} \right] 
           \left[ \frac {\lambda} {1100~{\rm \AA}} \right]  \;.
\end{eqnarray}
Here, we converted flux, $F_{\lambda}$, to luminosity, $L_{\lambda} = 4 \pi d_L^2 F_{\lambda}$,
using the luminosity distance, $d_L(z)$, computed for a flat $\Lambda$CDM universe with 
$H_0 = 70$ km~s$^{-1}$~Mpc$^{-1}$ and density parameters $\Omega_m = 0.275$ and 
$\Omega_{\Lambda} =0.725$ (Komatsu \etal\ 2011).
Table 3 shows the AGN indices from composite spectra found from various surveys (1991-2004) 
compared to the current COS survey (2012).  The median values of $\alpha_{\nu}$ from these surveys 
depend on the rest wavelengths sampled, since many quasar SEDs involve both power-law distributions
and broad features (``big blue bump", ``small UV bump") attributed to thermal emission from the accretion
disk (Shields 1978) and to blends of \FeII\ emission (Wills \etal\ 1985).  In general, the SEDs steepen from the 
optical and near-UV bands, where $F_{\nu} \propto \nu^{-0.3}$ or $\nu^{-0.4}$, to distributions 
$\nu^{-1.4}$ or $\nu^{-2.0}$ at shorter (FUV and EUV) wavelengths.  The break in the COS composite 
occurs at $\sim$1000~\AA, well shortward of  \Lya\ (1216~\AA).   In future \HST/COS archival studies, we 
intend to find AGN in the critical range $0.15 < z < 0.5$, whose rest-frame spectra probe the 
1000~\AA\ break.  

Figure 2  shows two outstanding examples of far-UV spectra taken by COS, labeling prominent  emission lines 
in the low-redshift Seyfert 1.5 galaxies,  Mrk 817 and Mrk~1513.    According to the classification scheme proposed
by Richards \etal\ (2011), Mrk 817 is a ``wind-dominated AGN" with weak \Lya, \SiIV, and \HeII\  
emission lines and blue-shifted \NV\ and \OIV].    Mrk~1513 is a narrow-line Seyfert galaxy, exhibiting a 
``disk-dominated" spectrum with prominent emission lines (\Lya, \CIV, \SiIV).   Figures 3--5 show individual 
spectra of all 22 AGN in our sample.   Table 4 lists the detected or expected emission lines and their rest wavelengths.  
Most of the low- and intermediate-redshift spectra show the prominent ultraviolet emission lines
seen in optical spectra of higher-redshift QSOs:  \Lya,  \CIV,  \NV, \OVI, \SiIV+\OIV].   
The weaker lines (centroids of blends) include \OI\ 1303 + \SiII\ 1307, \CII\ 1335, \HeII\ 1640, \OIII] 1665, and 
many others at shorter wavelengths.  A number of weak lines show up in the composite spectrum, 
including \FeIII\ 1123, \NII\ 1085, \NIII\ 989-991,  \NIII] 1750,  \NIV] 1486, and two emission lines of \CIII\  
(1175~\AA\ and 977~\AA).   The semi-forbidden line of \CIII] $\lambda1909$ lies outside the range of the 
COS G160M grating.

Figure 6 shows the composite UV/EUV spectrum for the full sample of 22 AGN, extending from 
550~\AA\ to 1750~\AA \ in the AGN rest frame.   Because these AGN range in redshift from 
$z_{\rm min}  = 0.026$ to $z_{\rm max} = 1.44$,  wavelengths in the composite include contributions 
from a limited number of spectra, typically 3--12,  as shown in the bottom panel  of Figure~6.  These 
effects are further illustrated in Figure~7, with composite spectra for subsets of seven 
low-redshift AGN and eight high-redshift AGN.  These subsets probe the FUV and EUV portions, 
respectively, in the AGN rest frame.  In the EUV range, we see a number of broad emission lines (Table 4)  
from higher ionization states of oxygen, nitrogen, and neon.  The most prominent EUV  emission-line blends 
are  \OII\ (834~\AA) and \OIII\ (833~\AA),  the \NeVIII\ doublet (770~\AA\ and 780~\AA), and \NeV\ (569~\AA).  
We also detect weaker lines of  \NIII\ (685~\AA),  \NIV\ (765~\AA), and a range of high ionization states of oxygen, 
including  \OIII\ (703~\AA), \OIV\ (788, 608, 554~\AA), and \OV\ (630~\AA).   Some of the difference in the \FUSE\
composite index (Scott \etal\ 2004) may come from incorrectly fitting the continuum through these EUV emission 
lines. These  lines from high ionization states provide important diagnostics of BELR physical conditions, 
particularly multiple ions from the same element such as \NeV\ and \NeVIII, or \OIII, \OIV, \OV, and \OVI.


\begin{figure*}
\epsscale{1.0}
\plotone{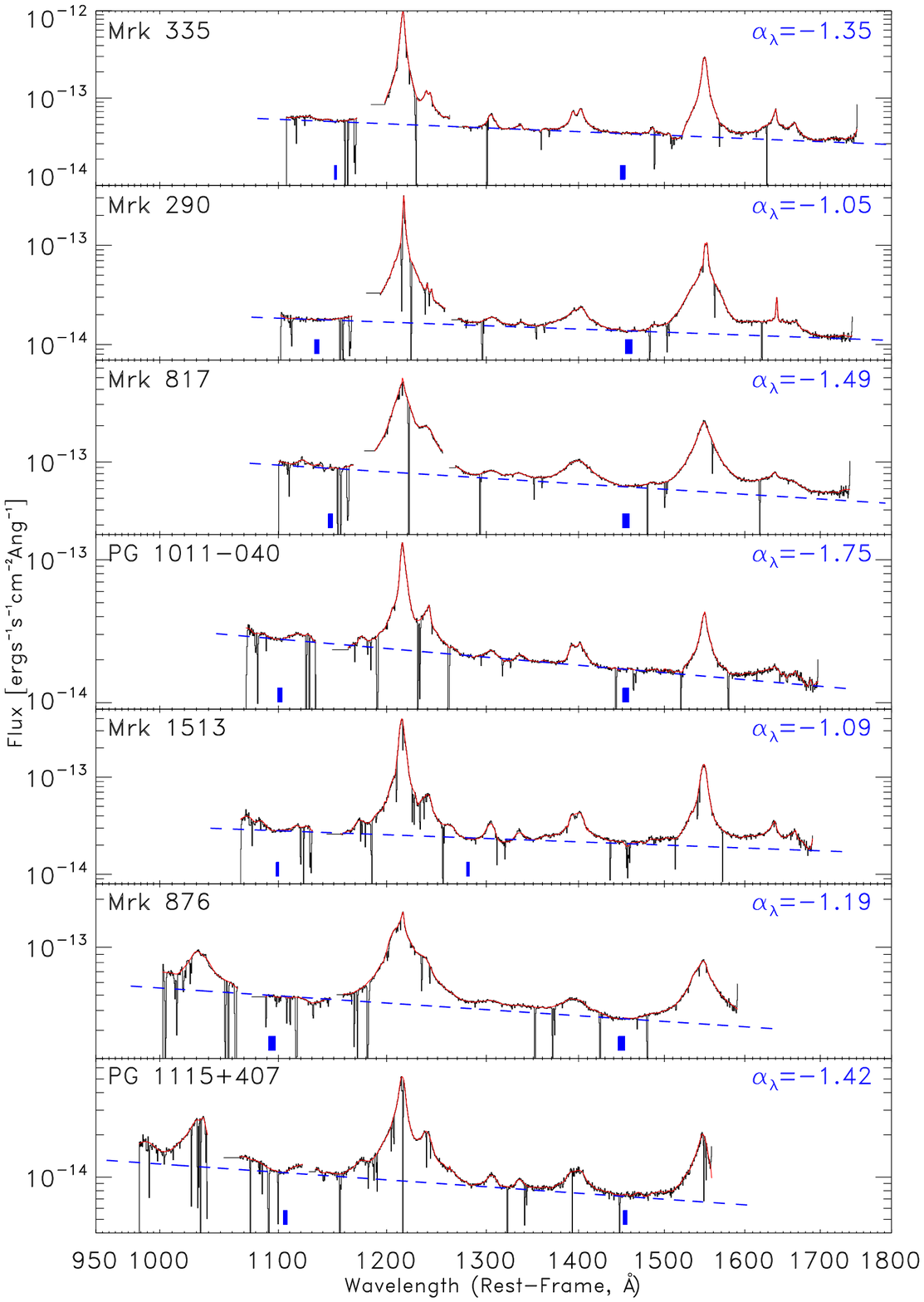}
\caption{First of three plots that show de-reddened  COS spectra and fitted spectral indices
$\alpha_{\lambda}$ (top right corners)  for seven low-redshift AGN ($0.026 \leq z \leq 0.154$) in the survey.   
Spectra are smoothed over 7 pixels and binned to 52 pixels (0.5~\AA).  Red line is a spline fit.  
} 
\end{figure*}



\begin{figure*}
\epsscale{1.0}
\plotone{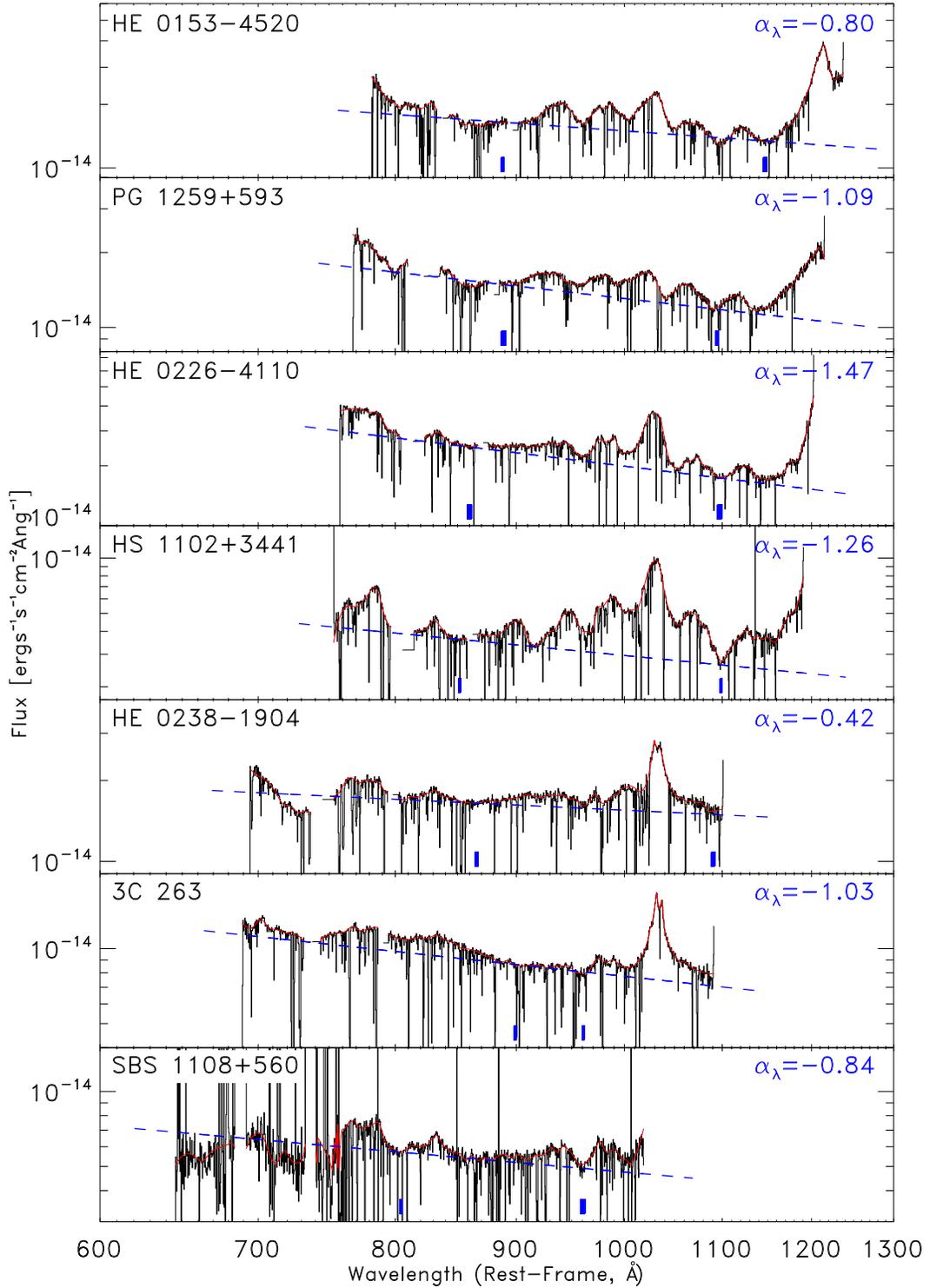}
\caption{Same as Figure 3, showing de-reddened COS spectra for seven intermediate-redshift AGN  
($0.45 \leq z \leq 0.77$) in survey.   Restored continuum  at $\lambda < 755$~\AA\ (AGN rest frame) 
toward SBS\,1108+560 is unreliable, owing to the presence of a strong LLS at $z = 0.46445$.  
 } 
\end{figure*}



\begin{figure*}
\epsscale{1.0}
\plotone{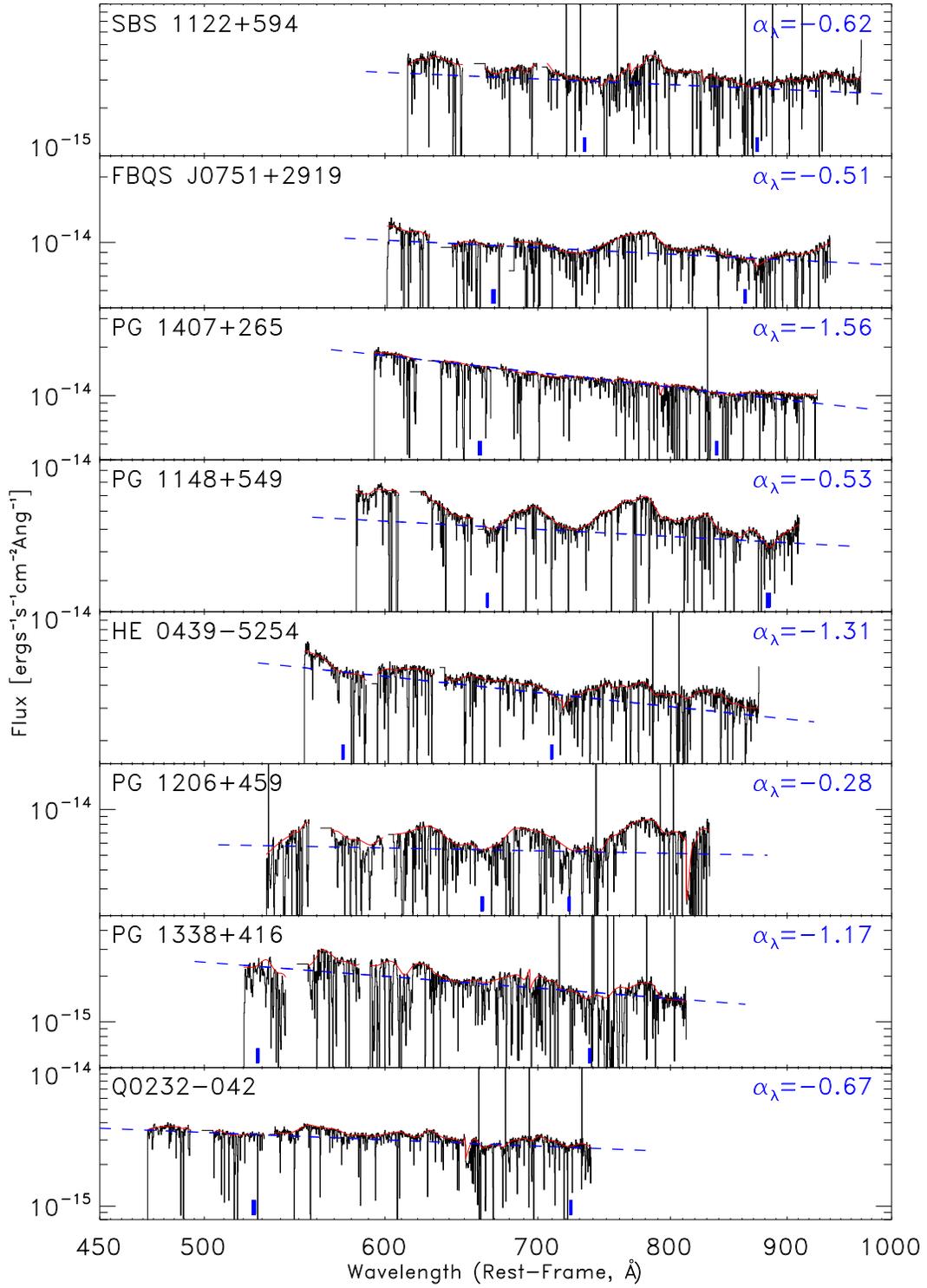}
\caption{Same as Figure 3, showing de-reddened COS spectra for eight high-redshift AGN ($0.85 \leq z \leq 1.44$)  
in the survey. 
} 
\end{figure*}



\begin{figure*}
\includegraphics[angle=90,scale=0.72]{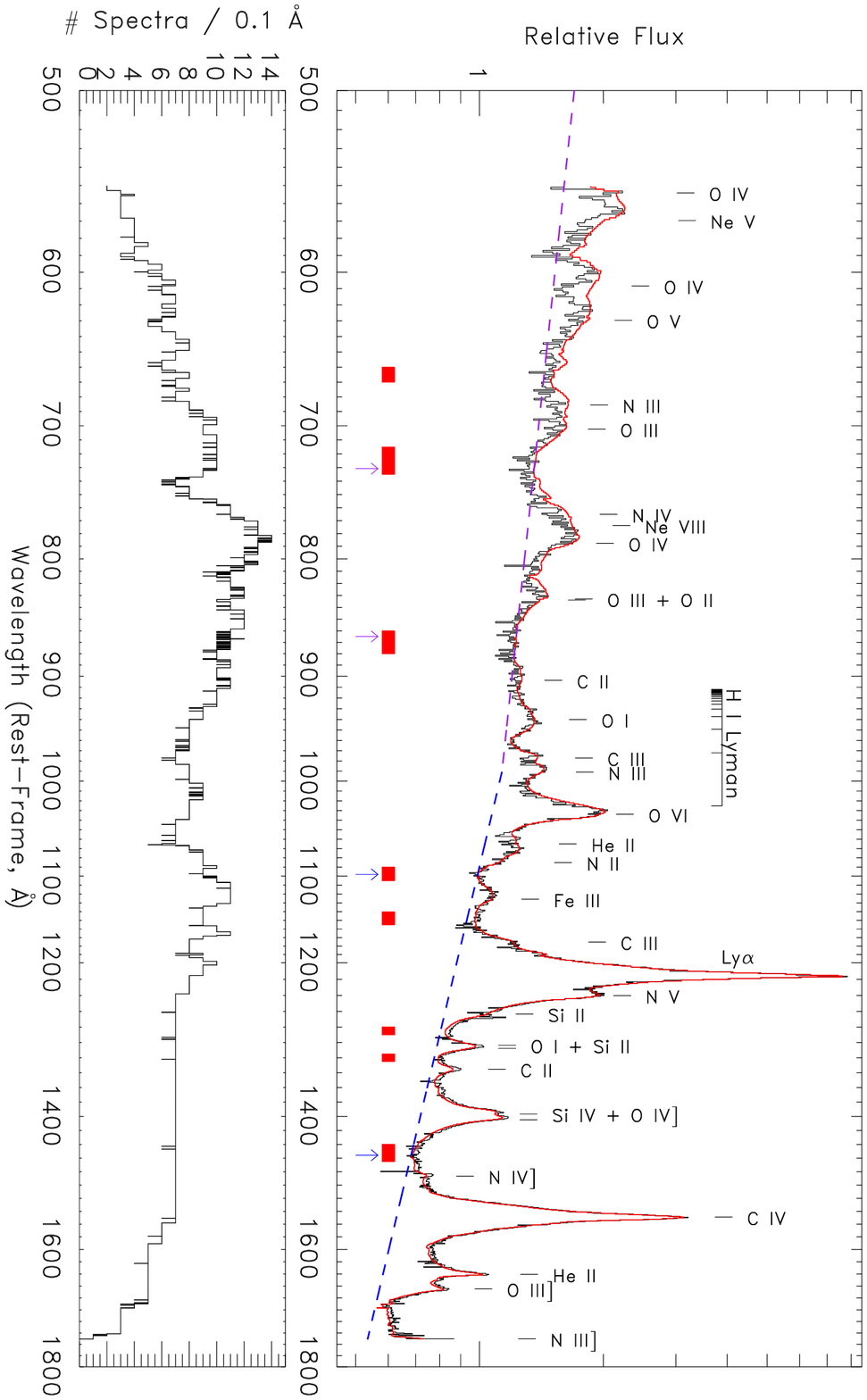}
\caption{Composite \HST/COS spectrum of all 22 AGN at $0.026 \leq z_{\rm em} \leq 1.44$, resampled to 0.1~\AA, 
plotted in 1~\AA\ bins, normalized to 1 at 1100 \AA, and showing array of broad FUV and EUV emission lines 
(550~\AA\ $< \lambda_{\rm rest} < 1220$~\AA) atop a power-law continuum.   Eight continuum windows are shown as
small red boxes.  Composite data are shown in black, and red curve is composite of individual spline fits.  Lower panel 
shows number of AGN targets contributing to composite spectrum vs.\ wavelength.   We can distinguish regions of true 
continuum (noted by small arrows) from emission lines 
and  ``pseudo-continuum'' composed of blended emission lines.  The power-law continuum exhibits a break at 
$\lambda \approx 1000\pm50$~\AA, with a flatter (softer) spectrum at EUV wavelengths.  The frequency distribution, 
$F_{\nu} \propto \nu^{\alpha_{\nu}}$, has indices
$\alpha_{\nu} = -1.41 \pm 0.21$ (EUV, $\lambda < 1000$~\AA) and
$\alpha_{\nu} = -0.68 \pm 0.14$ (FUV, $\lambda  >1200$~\AA).  
}  
\end{figure*}


Figure 8 looks for possible dependences of the spectral indices on AGN redshift and luminosity,
$\lambda L_{\lambda}$ (at 1100~\AA).    The symbols correspond to the redshift sub-groups (Table~1), 
which have mean values, $\langle \log \lambda L_{\lambda} \rangle =$ 44.44, 46.08, and 46.31 for groups at 
$\langle z \rangle$ = 0.070, 0.565, and 1.07, respectively.  We find a weak correlation of individual spectral 
indices with each.  Both likely reflect the same trend, in which higher-luminosity, higher-redshift AGN sample 
the rest-frame continuum at progressively shorter wavelengths.  We see a steepening 
of the flux distribution, $F_{\nu}$, which corresponds to a flattening in $F_{\lambda}$.  Because different 
redshift ranges sample the SED on both sides of the $1000$~\AA\ break, samples of AGN at higher 
redshifts are biased toward softer spectra.  These correlations will be explored further with
larger samples from COS archival data.   We will also explore possible dependence on black hole mass. 
Three of the AGN in our survey have black hole mass estimates obtained from observed \CIV\ line widths 
and continuum luminosities at 1350~\AA\ (Vestergaard \& Peterson 2006).   These mass estimates are:  
$\log (M_{\rm BH} / M_{\odot}) = 7.41$, 7.93, and 8.85 for Mrk~335, Mrk~817, and Mrk~876, respectively, 
based on UV data from \IUE\ and \HST/FOS.   Our new spectra will provide accurate \CIV\ line widths and 
1300~\AA\ luminosities for the first seven AGN in Table~1, at $z < 0.155$.  At higher redshifts, the \CIV\ emission  
line shifts out of the COS/G160M band.    The correlation between $M_{\rm BH}$ and \CIV\ line width 
is controversial (Netzer \etal\ 2007;  Assef \etal\ 2011) owing to scatter in the relation and likely changes
with AGN luminosity (Shen \etal\ 2008).    We will further analyze the AGN in our survey for correlations 
with \CIV\ and other emission lines.

\newpage


\begin{figure*}
\includegraphics[angle=90,scale=0.72]{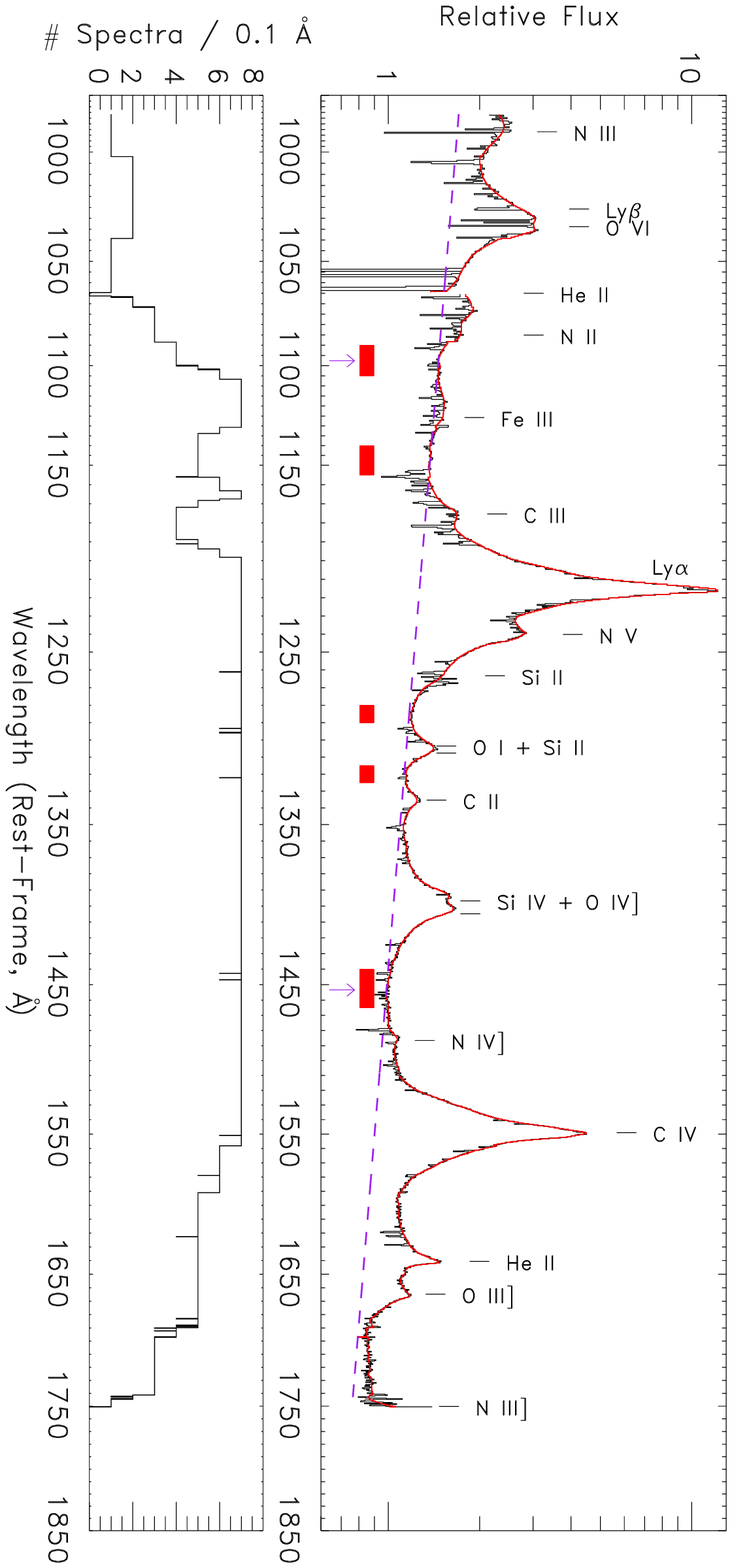}
\includegraphics[angle=90,scale=0.72]{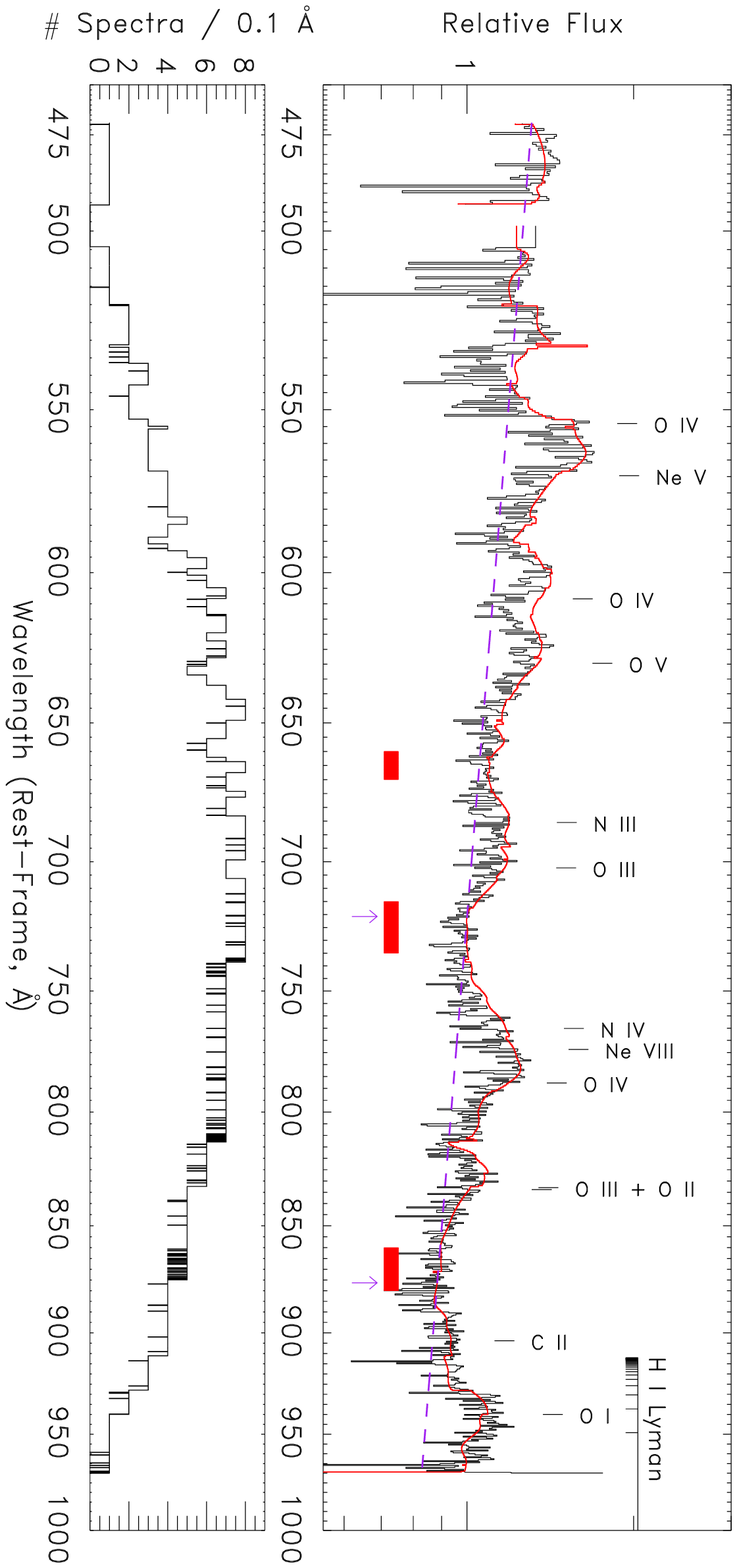}
\caption{ (Top) Composite COS spectrum of seven low-redshift AGN at $z_{\rm em} \leq 0.154$,  normalized at 1450 \AA, 
resampled to 0.1~\AA\ and plotted in 0.5~\AA\ bins. Continuum windows are shown as small red boxes, and red curve is 
our spline fit.   The FUV power-law continuum ($\lambda > 1200$~\AA) has index $\alpha_{\nu} = -0.58\pm 0.2$, similar 
to the overall (22-QSO) composite value of $-0.68 \pm 0.14$ shown in Figure 6.   
(Bottom)   Composite COS spectrum of eight high-redshift AGN at $0.85 < z_{\rm em} < 1.44$.
normalized at 733 \AA.   The EUV power-law continuum ($\lambda < 1000$~\AA) has index $\alpha_{\nu} = -1.37 \pm 0.20$, 
similar to the overall composite value of $-1.41 \pm 0.21$ in Figure 6.   
 }  
\end{figure*}


\clearpage


\begin{figure}
\epsscale{1.25}
\plotone{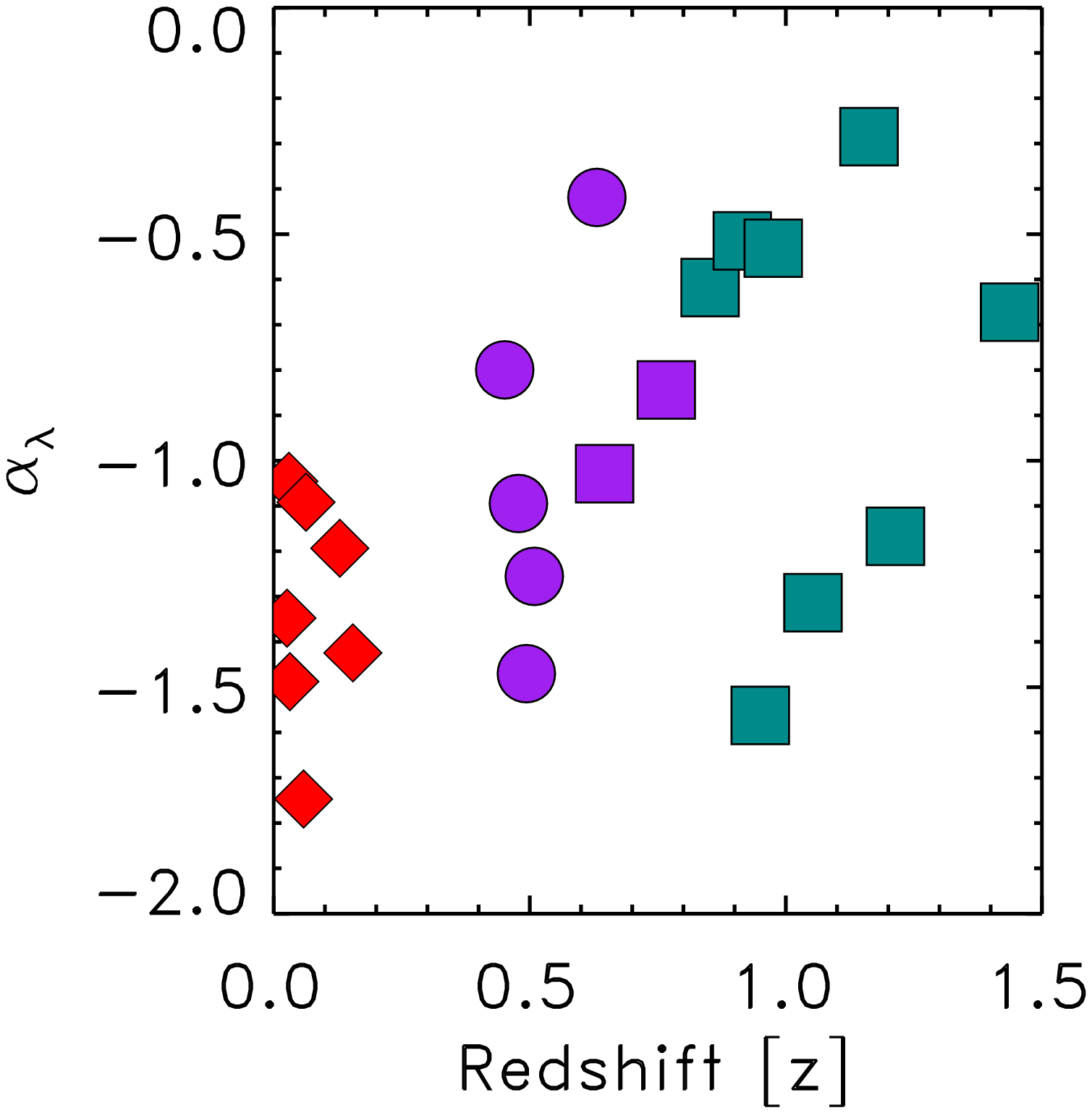} 
\plotone{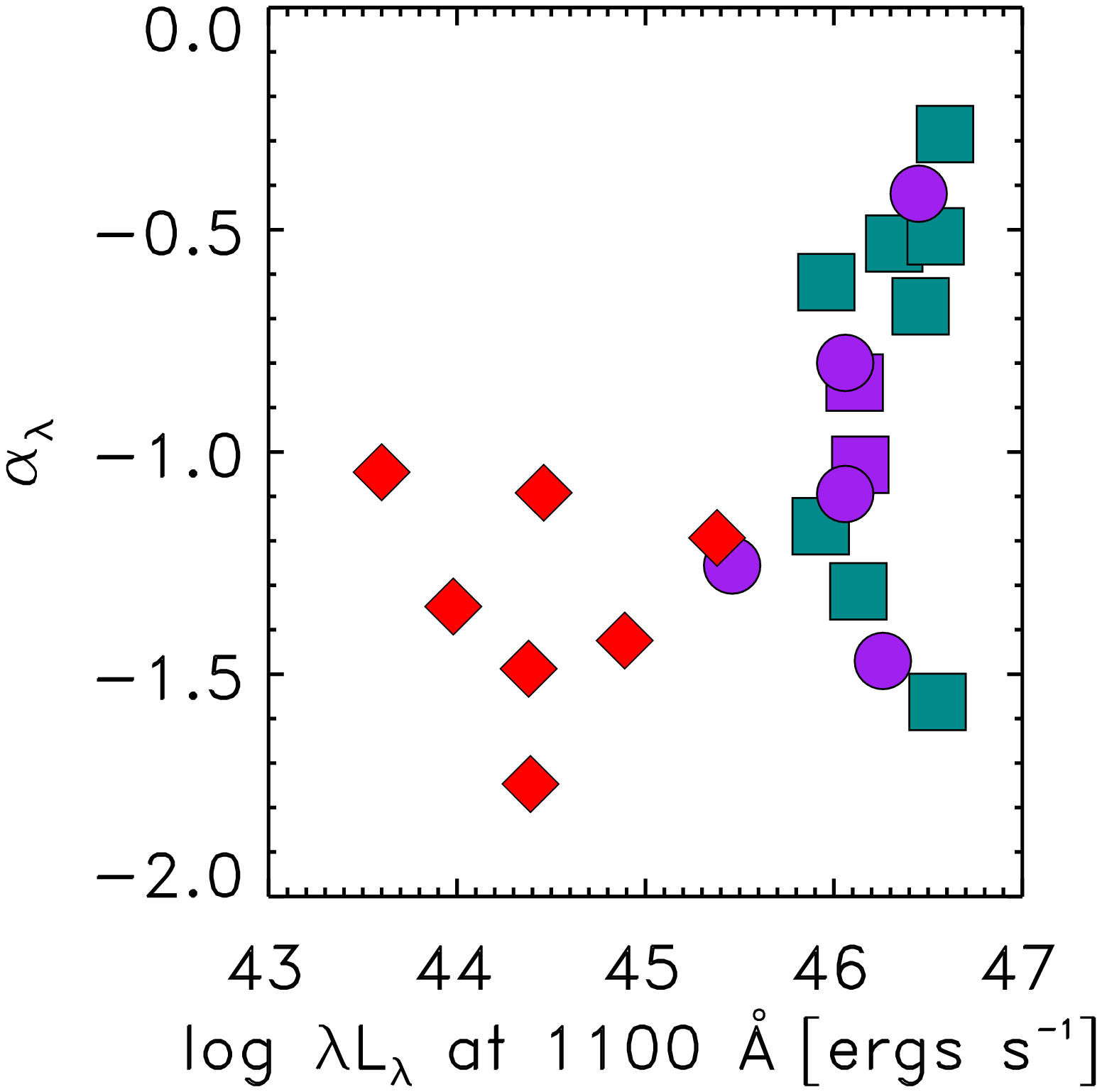}
\caption{{\small Correlations of fitted spectral indices, $\alpha_{\lambda}$, with AGN redshift (top) and 
1100~\AA\ luminosity (bottom) for our 22 AGN.  Colors correspond to the three redshift groups:
red (seven low-redshift AGN in Figure 3);  purple (seven intermediate-redshift AGN in Figure 4); 
turquoise (eight high-redshift AGN in Figure 5).   Symbol shapes refer to the wavelength windows
used to fit the continua:  diamonds have windows longward of 1000~\AA, squares have windows
shortward of 1000~\AA, and circles have windows that straddle 1000~\AA, with one window (each) 
in EUV and FUV. }
} 
\end{figure}



\begin{figure*}
\includegraphics[angle=90,scale=0.58]{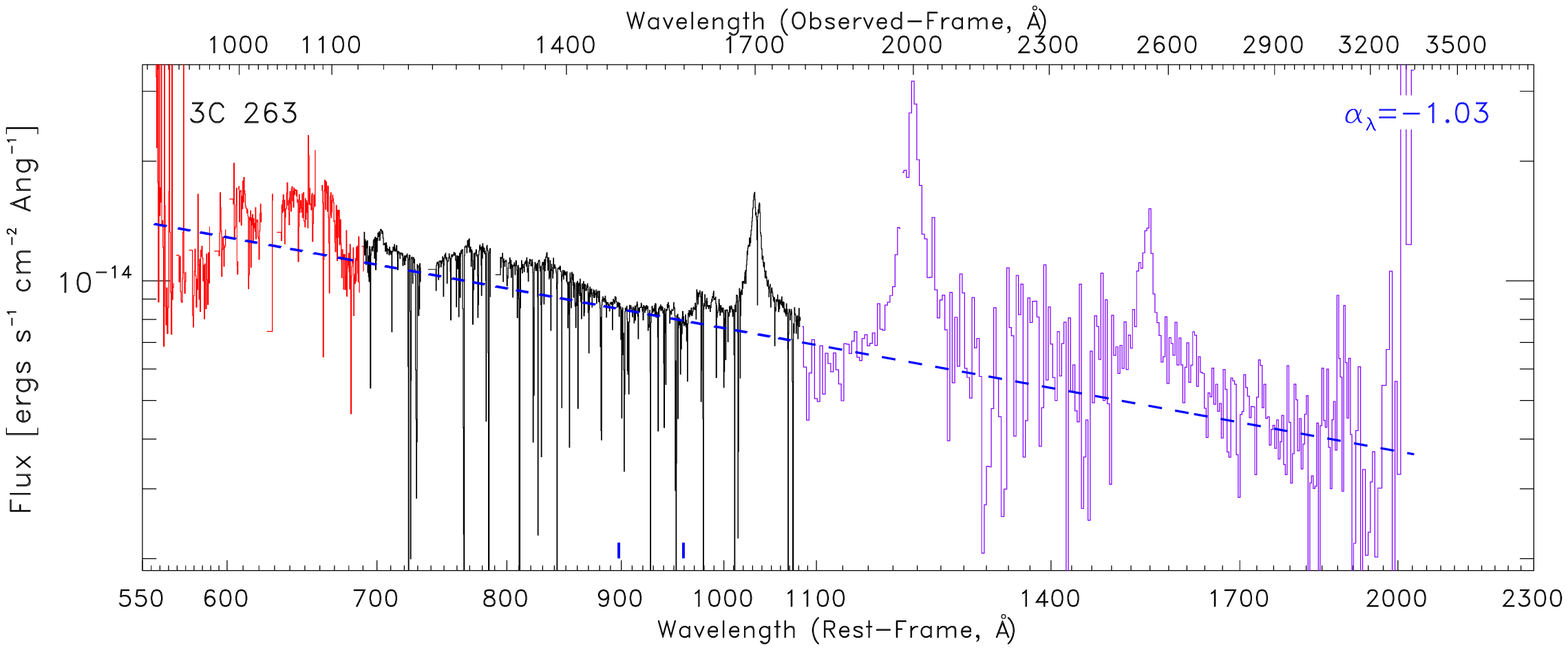}
\includegraphics[angle=90,scale=0.58]{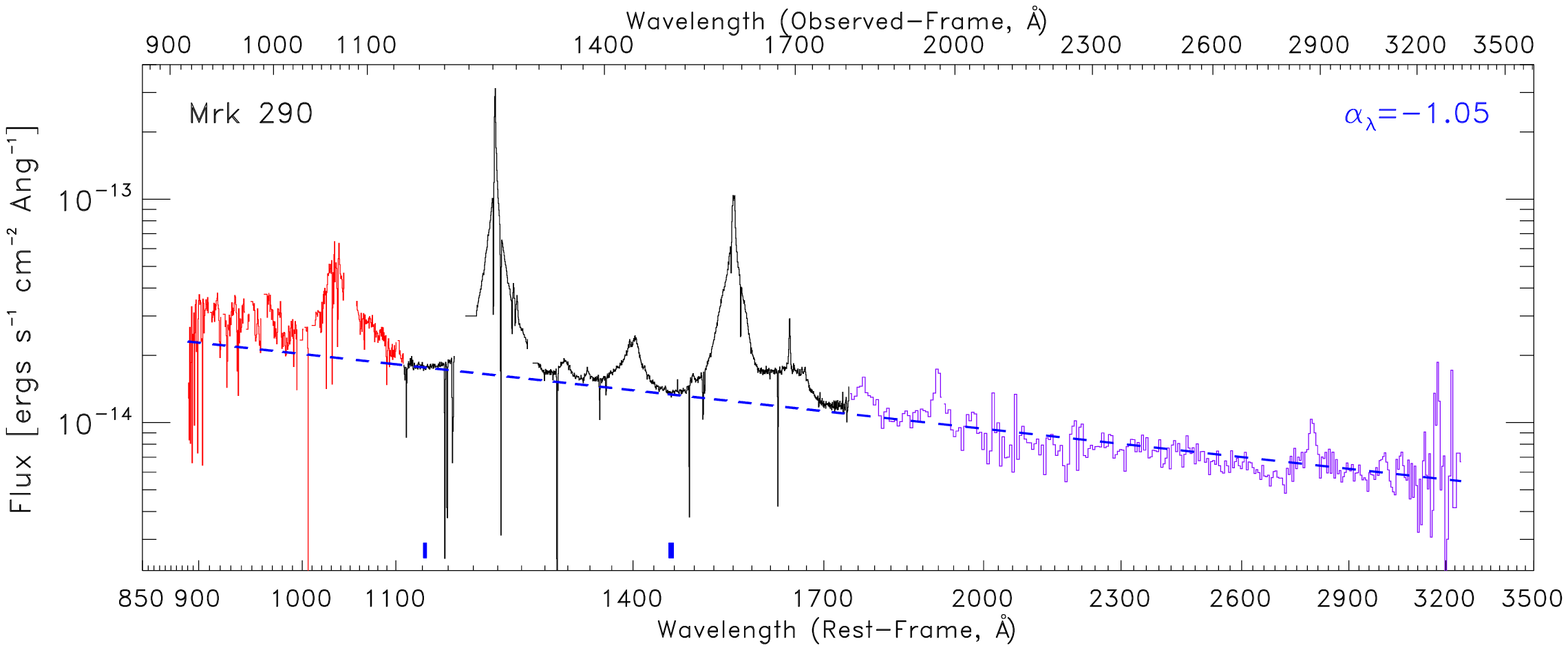}
\includegraphics[angle=90,scale=0.58]{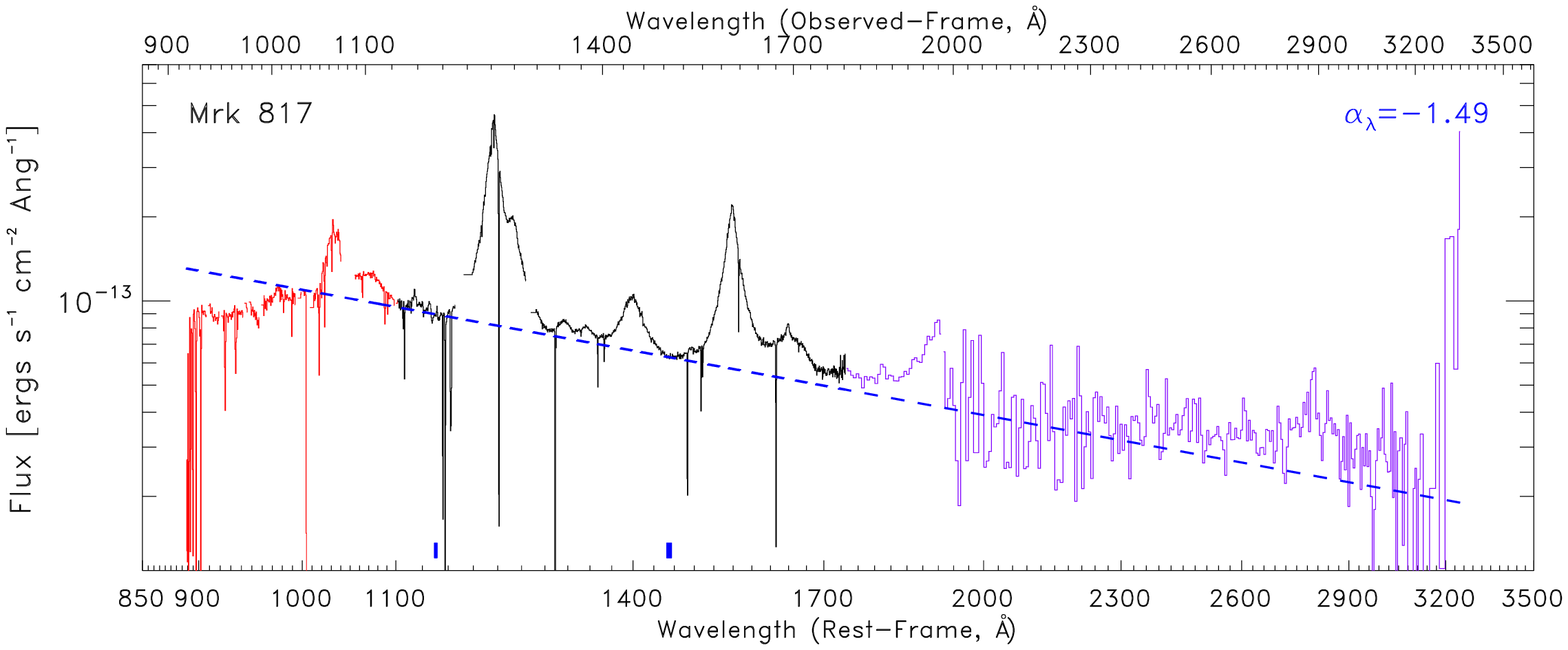}
\includegraphics[angle=90,scale=0.58]{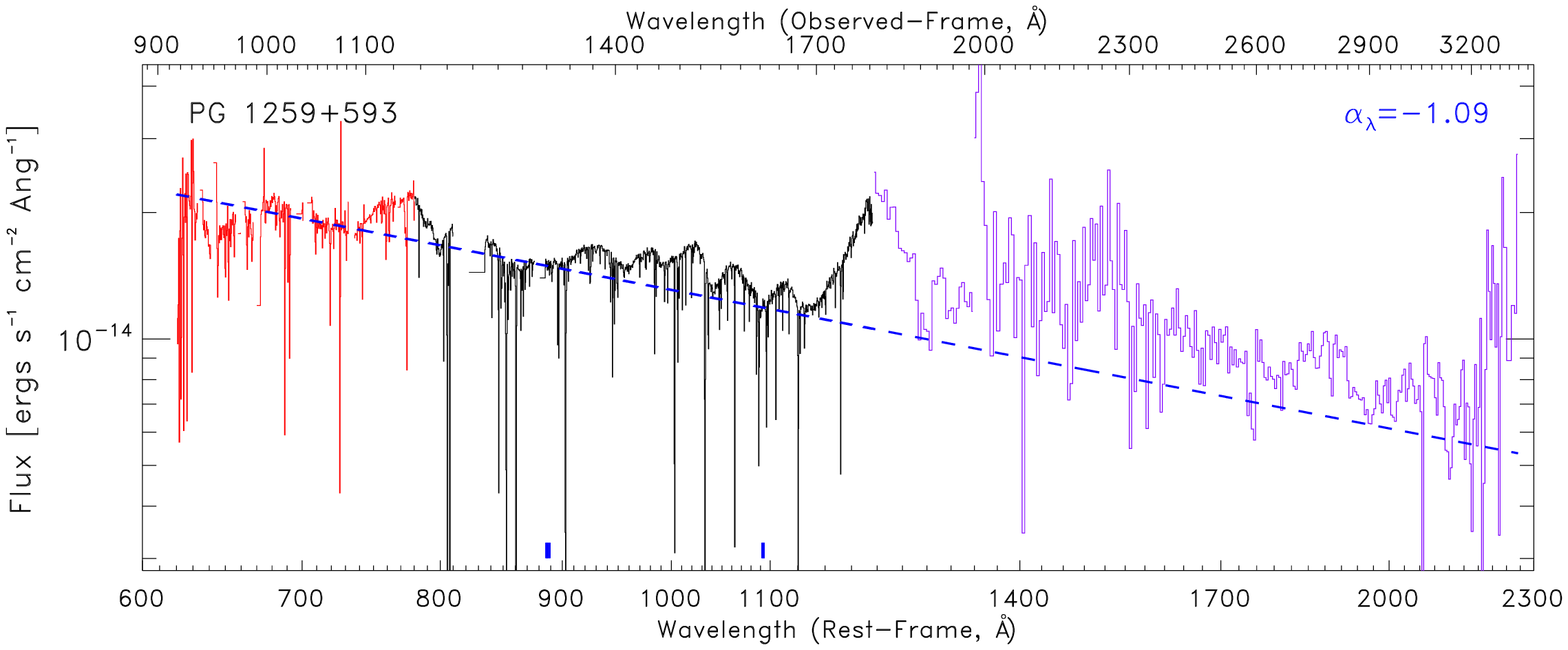}

\caption{ {\small Stitched spectra of 3C\,263, Mrk\,290, Mrk\,817, PG\,1259+593 (top to bottom)
 combining spectral data from \FUSE\ (shortest wavelengths), \HST/COS (middle range), and \IUE\
 (longest wavelengths).  Transitions are apparent from line texture. We excised \HI\ and \OI\ 
 airglow lines.  Data from \FUSE\ and COS are binned to 0.5~\AA, and \IUE\ data to 4.5~\AA\ (short-wavelength) 
 or  5.0~\AA\  (long wavelength).  The fitted spectral indices, $\alpha_{\lambda}$, are those from COS alone 
 (Table~1), but they also provide reasonable fits to the \FUSE\ and \IUE\ portions. }
   }  
\end{figure*}



\begin{figure}
\epsscale{1.2}
\plotone{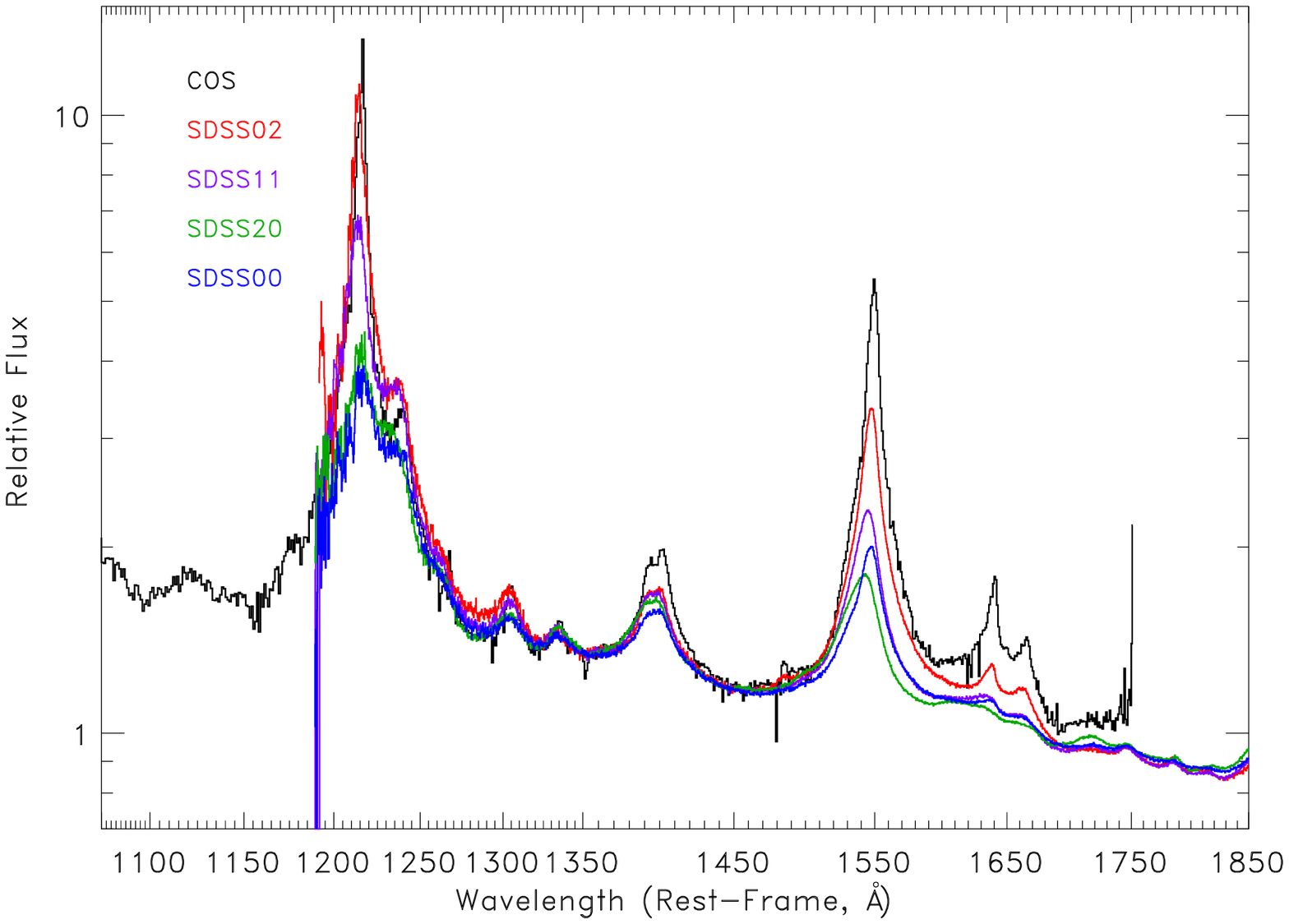}
\caption{Overplot of  COS composite spectrum with four composite spectra from SDSS (Richards \etal\ 2011;
   Hewett \& Wild 2010).   Four SDSS colors refer to AGN classes with progressively stronger emission lines, 
   referred to as  ``disk-type" (strong emission lines, hard spectra) and  ``wind-type" (weak emission lines, softer 
   spectra).  The hard spectra exhibit strong \HeII\ $\lambda1640$ emission lines.  Our 22 AGN  with COS 
   are primarily  ``disk type" with strong emission lines, particularly  \Lya, \CIV, \HeII, \OIII], \SiIV/\OIV]. 
} 
\end{figure}


Although the statistical quality of COS data exceeds that of \FUSE\ and \IUE, we thought it  prudent
to compare the spectra produced by these three ultraviolet spectrographs.  
Figure 9 shows stitched spectra for four AGN, with combined data from COS, \FUSE, and \IUE.   
The \FUSE\ and COS data were patched together at the following wavelengths:  Mrk~290 (1141.3~\AA), 
Mrk~817 (1136.6~\AA), PG\,1259+593 (1153.2~\AA), and 3C~363 (1136.6~\AA).  The COS and \IUE\ data
were patched together at 1796.5~\AA.   For these four AGN, the spectral indices fitted to \HST/COS data
provide reasonable fits to the data, both at shorter wavelengths (\FUSE) and longer wavelengths (\IUE).  

Figure 10 compares our COS composite spectrum with the AGN composites found from the Sloan
Digital Sky Survey (SDSS) using new redshifts from Hewett \& Wild (2010).  The SDSS composites
were most recently presented by Richards \etal\ (2011) in four sub-groups, defined by their 
line-to-continuum ratios.   The COS composite exhibits stronger emission lines than those of any of the 
SDSS sub-groups, suggesting that many of our AGN have ``disk-type" spectra defined by 
Richards \etal\ (2011) and more likely to be radio-loud with hard ionizing spectra.  However, 
only 2 of the 22 AGN (3C~263 and Q0232-042) are radio-loud, as measured by the ratio of fluxes 
at 5 GHz and optical (V band).   This 10\% fraction is consistent with the QSO radio-loud fraction.

We conclude this section with a few general results from the composite spectra.  We see no trace of 
the Lyman continuum edge  (912~\AA) in the rest frame of the composite spectrum, with an
optical-depth limit $\tau_{\rm HI} < 0.03$.  Possible explanations of the lack of a Lyman edge in
accretion-disk atmosphere models (Davis \etal\ 2007) include hot, highly ionized disk photospheres
and possible temperature inversions arising from external irradiation (Siemiginowska \etal\ 1995).  
We also see no obvious emission line from \HeII\ $\lambda584$, which could be present in the broad 
emission-line region.   Because this line is formed deep inside BELR clouds,  it will scatter 
resonantly and be destroyed by \HI\  Lyman continuum absorption.  However, only four AGN contribute 
to the composite at 584~\AA, and any weak \HeI\  line would appear in the broad red wings of
 \OIV\ $\lambda 554$ and \NeV\ $\lambda 569$.   We also see no ``Lyman Valley" (M\o ller \& Jakobsen 1990),
 an extremely broad absorption feature centered at $\lambda\sim650$~\AA\ that may arise from the cumulative 
 Lyman continuum edges of translucent \HI\ absorbers at moderate redshift.  
 The expected effect is small ($\sim5$\%) at $z<1$.

\newpage

\section{DISCUSSION AND CONCLUSIONS}

We now summarize the results and implications of our survey of AGN spectral distributions in the rest-frame 
FUV and EUV, using spectra of 22 AGN studied by \HST/COS.   We achieved the primary goal of our initial study, 
which was to investigate the difference between the markedly different spectral indices, $\alpha_{\nu}$, obtained 
by earlier surveys with \HST/FOS (Telfer \etal\ 2002) and \FUSE\ (Scott \etal\ 2004).    Our major results are: 
\begin{enumerate}
 
 \item Two previous ultraviolet surveys of the spectra of AGN gave conflicting results for their
    composite rest-frame LyC spectra taken by \HST/FOS and \FUSE.   High-S/N spectra of 22 
    AGN taken with \HST/COS G130M and G160M gratings give a composite EUV spectral index, 
    $\alpha_{\nu}  = -1.41 \pm 0.21$, in reasonable agreement with the HST/FOS value,
    $\alpha_{\nu}   = -1.57\pm0.17$ (Telfer \etal\ 2002) for radio-quiet AGN.  
    
\item For wavelengths $\lambda > 1200$~\AA, previous AGN surveys found spectral indices $\alpha_{\nu}$ 
   between $-0.3$ and $-0.6$, steepening to $\alpha_{\nu} \approx -1.6$ at $\lambda < 1000$~\AA.  
   The composite indices depend on sample size and the range of QSO luminosity and redshift.  The
   continuum must be carefully fitted beneath strong EUV emission lines and is sensitive to how 
   one combines QSOs with a wide range of individual spectral indices.  
  
\item The 22 AGN exhibit a wide range of individual AGN spectral indices at $\lambda < 1100$~\AA, 
    ranging from $\alpha_{\nu} = -0.25$  (for PG~1011-040 with a flat, hard spectrum) to 
    $\alpha_{\nu} = -1.75$ (for PG~1206+459 with a steep, soft spectrum).   Most spectra show a break 
    at rest wavelength $\lambda \approx 1000$~\AA,  with a steepening of the flux distribution, 
    $F_{\nu} \propto \nu^{\alpha_{\nu}}$, at higher frequencies and $\lambda < 1000$~\AA.

\item We see no Lyman edge at 912~\AA\ or \HeI\ $\lambda584$ emission line in the AGN composite, 
  which has implications for models of accretion-disk atmospheres and AGN emission-line regions.  Our 
  COS AGN composite has stronger emission lines (\Lya, \CIV, \SiIV, \HeII) than the SDSS composites
  (Richards \etal\ 2011), suggesting that these AGN are of ``disk type", with hard ionizing spectra and
  strong \HeII\ $\lambda1640$ emission.

\end{enumerate}


\acknowledgments

\noindent
We thank Brian Keeney,  St\'ephane B\'eland, and the COS/GTO team for help on the calibration 
and verification of  early COS data, and Gordon Richards and Paul Hewett for sharing their SDSS composite
spectra.   Brad Peterson and Paulina Lira provided advice on the use of \CIV\ as a black-hole
mass diagnostic. John Stocke, Shane Davis, and James Green provided helpful comments on this project.  
This research was supported by NASA grants NNX08-AC14G and NAS5-98043 and the Astrophysical Theory Program 
(NNX07-AG77G from NASA and AST07-07474 from NSF) at the University of Colorado Boulder.


\clearpage


\begin{deluxetable}{lllcl}
\tabletypesize{\footnotesize}
\tablecaption{AGN Composite Spectral-Index Fits (FUV and EUV)}
\tablecolumns{5}
\tablewidth{0pt}
\tablehead{
  \colhead{Survey Reference\tablenotemark{a}}       &
  \colhead{$N_{\rm QSO}$\tablenotemark{a}}            &
  \colhead{$\alpha_{\nu}$\tablenotemark{a}}             &
  \colhead{$\lambda$-range\tablenotemark{a}}        &
  \colhead{Survey Characteristics\tablenotemark{a}}  
            }
\startdata
Francis \etal\ (1991)           &    718  &  $-0.32$                &   $1450-5050$~\AA\  &  LBQS (Large Bright Quasar Survey)  \\
Schneider \etal\ (1991)      &      30  &  $-0.92\pm0.05$                &   $1270-1600$~\AA\  &   CCD grism survey ($z > 3.1$)  \\
Carballo \etal\ (1999)         &     48   &  $-0.87\pm 0.20$ &   $1300-4500$~\AA\  &   B3-VLA (radio) QSOs  ($z < 1.2$)  \\
Carballo \etal\ (1999)         &     48   &  $-0.40\pm 0.14$ &   $1300-4500$~\AA\  &   B3-VLA(radio)  QSOs  ($z >1.2$)  \\
Vanden Berk \etal\ (2001) &   2200 & $-0.44$                 &   $1300-5000$~\AA\  &  SDSS ($z = 0.044-4.78$) \\
Brotherton \etal\ (2001)      &    657  & $-0.46$                 &   $1220-4800$~\AA\  &  FBQS (First Bright Quasar Survey)  \\ 
                                                &             &                                &                                       &            \\
Telfer \etal\ (2002)              &     77   &  $-1.57\pm0.17$  &   $500-1200$~\AA\    &  \HST/FOS (radio-quiet sample)  \\
Telfer \etal\ (2002)              &   107   &  $-1.96\pm0.12$  &   $500-1200$~\AA\    &  \HST/FOS (radio-loud sample)  \\
Scott \etal\ (2004)               &      85   &  $-0.56^{+0.38}_{-0.28}$&   $630-1155$~\AA     & \FUSE\ ultraviolet ($z < 0.67$)   \\
This paper                           &      15   &  $-1.41\pm 0.21$  &   $550-1000$~\AA\    & \HST/COS ultraviolet ($0.45 < z < 1.44$) \\
This paper                           &      7     &  $-0.68\pm 0.14$  &   $1200-1750$~\AA\    & \HST/COS ultraviolet ($z < 0.16$) \\
 \enddata

\tablenotetext{a}{Top group of surveys sample visible/FUV wavelengths ($\lambda > 1200$~\AA), 
    while lower group samples EUV ($\lambda < 1000$~\AA).
    The columns list:  composite survey reference;   mean or median spectral index; 
   number of QSOs in survey;  rest-wavelength range of fit;  survey characteristics.}
 
\end{deluxetable}



\begin{deluxetable}{lrl}
\tabletypesize{\footnotesize}
\tablecaption{AGN Emission Lines and Blends (FUV and EUV)}  
\tablecolumns{3}
\tablewidth{0pt}
\tablehead{   \colhead{Ion}   &   \colhead{Wavelength\tablenotemark{a}}   &   \colhead{Comments\tablenotemark{a} } }
\startdata
\NIII]       &   1750.2                  &  Semi-forbidden lines [2s\,2p$^2 \, (^4$P) $\rightarrow$ 2s$^2$\,2p$ \,(^2$P)] (blend of 1749.7, 1752.2)  \\
\OIII]       &   1664.7                  &  Semi-forbidden lines [2s\,2p$^3 \, (^5$S) $\rightarrow$ 2s$^2$\,2p$^2 \,(^3$P)] (blend of 1666.2, 1660.8) \\
\HeII\     &   1640.5                   &  \HeII\ (Balmer-$\alpha$) line    \\
\CIV\     &    1549.0                  &   Permitted doublet  [2p\,($^2$P) $\rightarrow$ 2s\,$(^2$S)]  (blend of 1548.19, 1550.77)  \\
\NIV]     &    1486.5                  &   Semi-forbidden line [2s\,2p\,$(^3$P$_1) \rightarrow$ 2s$^2\,(^1$S$_0)$]  \\
\OIV]     &    1404.8                  &   Semi-forbidden lines (blend of 1407.4, 1404.8, 1401.2) \\
\SiIV\    &    1396.75                &   Permitted doublet  [3p\,($^2$P)\,$\rightarrow$ 3s\,$(^2$S)]  (blend of 1393.76, 1402.77) \\
\CII\       &    1335.31               &  Three permitted lines (blend of 1334.43, 1335.66, 1335.71)  \\
\SiII\      &    1307.64               &  Permitted doublet (blend of 1304.37, 1309.28)  \\
\OI\        &    1303.49               &  Three permitted lines (blend of 1302.17, 1304.86, 1306.03) \\
\SiII\      &    1263.31               &  Three permitted lines (blend of 1260.42, 1264.74, 1265.00) \\
\NV\      &    1240.15               &   Permitted doublet  [2p\,($^2$P) $\rightarrow$ 2s\,$(^2$S)]  (blend of 1238.82, 1242.80) \\
\HI\       &    1215.67                &  Permitted \HI\ (Lyman-$\alpha$) line    \\
\CIII\      &   1175.25                &  Subordinate line [2p$^2 $\,$(^3$P)\,$\rightarrow$ 2s\,2p\,$(^3$P)] \\
\FeIII\    &   1125.79                &  Many lines (1122.53 -- 1131.91) from [3d$^5$\,4p\,($^5$P) $\rightarrow$ 3d$^6\,(^5$D)] \\
\NII\       &   1085.12                &  Many lines (1083.99 -- 1085.70) from  [2s\,2p$^3\, (^3$D) $\rightarrow$ 2s$^2$\,2p$^2 (^3$P)] \\
\OVI\     &   1033.82                &   Permitted doublet  [2p\,($^2$P) $\rightarrow$ 2s\,$(^2$S)] (blend of 1031.93, 1037.62,  \HI\ \Lyb\ 1025.72) \\
\NIII\      &    990.98                 &  Three permitted lines (blend of 989.80, 991.51, 991.58)   \\ 
              &                                  &                       \\
\CII\       &   903.8                    &  Two permitted lines  [2s\,2p$^2 \, (^2$P) $\rightarrow$ 2s$^2$ 2p$ \,(^2$P)]  \\
\OII\       &   833.8                    &  Three permitted lines [2s\,2p$^4 \, (^4$P) $\rightarrow$ 2s$^2$ 2p$^3\,(^4$S)]    \\
\OIII\      &   832.9                    &  Three permitted lines [2s\,2p$^3 \, (^3$D) $\rightarrow$ 2s$^2$ 2p$^2\,(^3$P)]    \\
\OIV\     &  787.7                     &  Three permitted lines [2s\,2p$^2 \, (^2$D) $\rightarrow$ 2s$^2$ 2p$\,(^2$P)]   \\
\NeVIII\ &  773.7                     &   Doublet  [2p\,($^2$P) $\rightarrow$ 2s\,$(^2$S)]   (blend of 770.41, 780.32)     \\
\NIV\      & 765.1                     &   Permitted line [2s\,2p$ \, (^1$P$_1$) $\rightarrow$ 2s$^2$\,($^1$S$_0$)]    \\
\OIII\       & 702.3                     &   Permitted lines  [2s\,2p$^3 \, (^3$P) $\rightarrow$ 2s$^2$ 2p$^2 \, (^3$P)] \\
\NIII\       & 685.5                     &   Permitted lines [2s\,2p$^2 \, (^2$P) $\rightarrow$ 2s$^2$ 2p\,$(^2$P)] \\
\OIV\      &  608.4                    &   Permitted lines   [2s\,2p$^2 \, (^2$S) $\rightarrow$ 2s$^2$\,2p$ \,(^2$P)]  \\
\OIV\      & 554.1                     &  Permitted lines  [2s\,2p$^2 \, (^2$P) $\rightarrow$ 2s$^2$\,2p$ \,(^2$P)]  \\
\NeV\     & 569.7                     & Permitted lines [2s\,2p$^3  \, (^3$D) $\rightarrow$ 2s$^2$\,2p$^2$\,($^3$P)]  \\
\OV\       & 629.7                     &  Permitted line  [2s\,2p ($^1$P$_1$) $\rightarrow$ 2s$^2 \, (^1$S$_0$)]   \\
\enddata

\tablenotetext{a}{Permitted and semi-forbidden lines, configurations, transitions, blends, and statistically 
weighted wavelengths.  Data sources:  Morton (2003), Verner \etal\ (1994), and NIST Data Tables.  }

\end{deluxetable}


\end{document}